\documentclass[aps,prd,twocolumn,superscriptaddress]{revtex4-2}
\pdfoutput=1

\usepackage{graphicx,bm,amsmath,amssymb}

\usepackage{hyperref}
\usepackage{xcolor}
\usepackage{mathtools}
\usepackage{mathrsfs}

\allowdisplaybreaks

\newcommand{\ud}{{\rm d}}
\newcommand{\gray}[1]{{\color{gray}{#1}}}
\newcommand{\0}{{\color{gray!30} 0}}

\begin{document}

\title{Scalable quantum computation of Quantum Electrodynamics beyond one spatial dimension}

\author{Zong-Gang Mou}%
\email{Z.G.Mou@soton.ac.uk}

\author{Bipasha Chakraborty}
\email{B.Chakraborty@soton.ac.uk}

\affiliation{%
School of Physics and Astronomy, University of Southampton, Highfield, Southampton, SO17 1BJ, United Kingdom
}%

\begin{abstract}

In the Hamiltonian formulation, Quantum Field Theory calculations scale exponentially with spatial volume, making real-time simulations intractable on classical computers and motivating quantum computation approaches. In Hamiltonian quantisation, bosonic fields introduce the additional challenge of an infinite-dimensional Hilbert space. We present a scalable quantum algorithm for Quantum Electrodynamics (QED), an Abelian gauge field theory in higher than one spatial dimensions, designed to address this limit while preserving gauge invariance. In our formulation, Gauss’s law is automatically satisfied when the implementation remains fully gauge invariant. We demonstrate how gauge invariance is maintained throughout the lattice discretisation, digitisation, and qubitisation procedures, and identify the most efficient representation for extending to large Hilbert space dimensions. Within this framework, we benchmark several quantum error mitigation techniques and find the calibration method to perform most effectively. The approach scales naturally to larger lattices, and we implement and test the 2+1 and 3+1 dimensional setups on current quantum hardware. Our results indicate that next-generation quantum platforms could enable reliable, fully quantum simulations of large-scale QED dynamics.

\end{abstract}

\maketitle

\section{Introduction}

\begin{figure*}[t!]
\includegraphics[width=\linewidth]{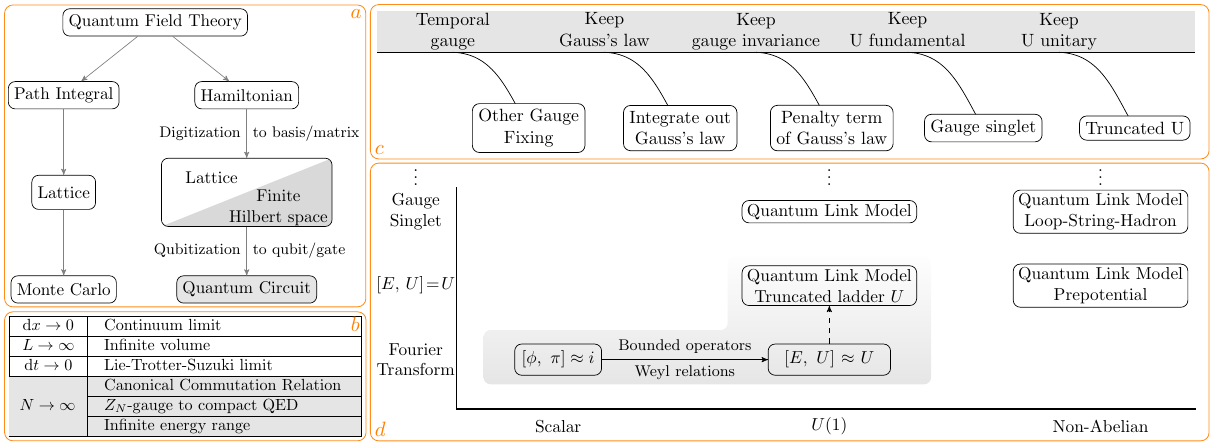}
\caption{(a) Hamiltonian quantisation requires an additional truncation of the infinite Hilbert space compared with the Path Integral.
(b) A unique $N\to\infty$ limit restores the full Hilbert space in the Hamiltonian approach.
(c) Main schemes for quantum simulation of Abelian gauge theory; the gray path indicates our economical, gauge-invariant method using a unitary (cyclic ladder) link and the Quantum Fourier Transform.
(d) The unitary treatment extends scalar-field algorithms to gauge fields, while ladder-based methods remain gauge invariant at large-$N$ and can generalise to non-Abelian theories.
}
\label{fig:intr}
\end{figure*}

A central goal of first-principles quantum field theory (QFT) simulation is to solve the Schr\"odinger functional equation in its general form\cite{Jackiw:1988sf}. Although QFT possesses unique features beyond quantum mechanics - such as regularisation and renormalisation arising from its infinite degrees of freedom, it is useful to view it as a collection of coupled quantum oscillators, each representing a field mode on a spatial lattice. In direct analogy to the Schr\"odinger equation of quantum mechanics, the QFT dynamics are governed by the Schr\"odinger functional equation\cite{Jackiw:1988sf}:
\begin{gather*}
i \frac{\partial}{\partial t} |\Psi \rangle = \hat H \left(\hat\Pi,\hat\phi\right) | \Psi\rangle 
,~
{\rm with}~
[\hat \phi,\hat\Pi]=i.
\end{gather*}
Here we can think of $\phi/\Pi$ either as $x/p$ in Quantum Mechanics, or as the scalar field $\phi(x)/\Pi(x)$ with the spatial index suppressed for simplicity.
For bosonic fields, however, the commutation relation requires an infinite-dimensional representation, which renders direct calculation impractical.

As a first-order differential equation in time, the Schr\"{o}dinger equation admits a formal solution in the shape of the path integral,
\begin{align*}
\Psi=e^{-i\hat Ht}\Psi_0=\int {\mathcal D}\phi \exp\left(i\int_{0}^{t}\ud t L\right) \Psi_0
.
\end{align*}

Treating the path integral as a distribution allows it to be interpreted as a statistical theory with a generally complex probability distribution function. Observables can then be obtained from samples generated according to this distribution. On a space-time lattice, the distribution becomes finite-dimensional, and results can be extrapolated to the continuum limit. The main challenge is not the dimensionality but the complex nature of the distribution. In special cases, such as imaginary time, the distribution becomes real and positive, enabling Markov Chain Monte Carlo methods which are the basis of lattice QCD \cite{Wilson:1974sk}. In general, however, the infamous sign problem arises when sampling from a complex distribution. Despite various proposed remedies for fermionic \cite{Chandrasekharan:1999cm} and bosonic \cite{Alexandru:2020wrj} systems, no universal solution exists in this case. This limitation underscores the need for quantum computers, which can inherently bypass the sign problem by representing quantum amplitudes directly.

In both continuum and lattice formulations, the Hamiltonian and path-integral approaches are equivalent as long as derivatives or infinite matrices are retained. Numerically, the Hamiltonian method requires truncation discretising derivatives or limiting Hilbert-space dimension. Therefore, the canonical commutation relations hold only approximately.
This extra approximation is absent in the path-integral formulation, where exact commutation relations are built in.

Once we attain the finite dimensional matrix equation as an approximation either from the differential equation or from the infinite dimensional matrix representation, we can further solve the equation by the matrix method.
Essentially, all the temporal evolution can be easily known once the eigenvalues and eigenvectors of the Hamilton are known,
\begin{align*}
\Psi=e^{-i\hat Ht}\Psi_0
={\mathscr U}^\dagger e^{-i\Lambda t}{\mathscr U}\Psi_0
,\quad
{\rm given}\quad
H={\mathscr U}^\dagger \Lambda {\mathscr U}
.
\end{align*}
The Exact Diagonalisation is in practice a challenging method for a large system.
The computational complexity of eigen-decomposition of a $N\times N$ matrix is ${\mathcal O}(N^3)$, and recall that in the Quantum Field Theory $N$ scales exponentially with the spatial volume.
Also, when the task is to find all eigenvalues, whether the matrix is sparse or not does not make a difference with the present matrix methods.
Instead of decomposing the matrix, we can reduce the computational complexity via the Lie-Trotter-Suzuki expansion,
\begin{align*}
\Psi=e^{-i\hat Ht}\Psi_0
=\lim_{\ud t\to 0} \left(e^{-i\hat H_{\Pi}\ud t} e^{-i\hat H_{\phi}\ud t}\right)^{\frac{t}{\ud t}} \Psi_0
.
\end{align*}

When each elementary matrix operation is efficiently executable, the total computational cost scales linearly with the matrix size and the number of Trotter steps, and accurate results can often be achieved with only a few steps.
Nevertheless, classical computation remains infeasible due to the exponential memory required to store and update the full wavefunctional of size $\sim N$. Quantum computers, by contrast, can perform these operations naturally through local single-qubit and two-qubit gates. The trade-off is that the wavefunctional cannot be inspected during evolution, and can only be measured at the end. Therefore, time evolution must be reconstructed from separate runs at different final times.

This work follows that lineage: starting from the Schr\"odinger functional equation, reducing it to a finite matrix form, and encoding it onto qubits. We choose the configuration-field basis because the QFT action is local in this representation; transforming to other bases introduces non-local interactions whose number grows with spatial volume.
Avoiding such non-locality is essential for studying the infinite-volume limit efficiently on quantum hardware.

The first quantum algorithm for scalar field theory was proposed in Ref.~\cite{Jordan:2012xnu}, following the same Hamiltonian-based prescription described above. Subsequent works~\cite{Somma:2015bcw,Macridin:2018gdw,Macridin:2018oli,Klco:2018zqz} refined its efficiency. As the scalar field is the simplest quantum field, it requires no fundamentally new algorithmic approach. Here we move to a more realistic system - Quantum Electrodynamics (QED), an Abelian gauge theory. Gauge theories introduce additional constraints, making the commutation relations harder to realise. This difficulty is amplified in non-Abelian cases, though we restrict our study to the Abelian case.

The 1+1-dimensional QED (Schwinger model) has been widely studied on quantum computers~\cite{Chakraborty:2020uhf,Shaw:2020udc,Zache:2021ggw,Angelides:2023noe,Farrell:2023fgd}. Its simplicity stems from the fact that the gauge field can be integrated out via Gauss’s law, leaving a fermion-only theory representable on qubits through the Jordan–Wigner or Bravyi–Kitaev mappings~\cite{Bravyi:2000vfj}. This simplification, however, fails in higher dimensions. Even after applying Gauss’s law, many gauge degrees of freedom remain, and the bosonic gauge field cannot be represented exactly in a finite Hilbert space without violating its commutation relations.

For 2+1-dimensional QED, several approaches~\cite{Haase:2020kaj,Paulson:2020zjd,Clemente:2022cka,Crippa:2024cqr,Bauer:2021gek,Kane:2022ejm,DAndrea:2023qnr,Burbano:2024uvn,Bender:2020ztu} attempt to eliminate Gauss’s law analogously to gauge fixing, which preserves simple canonical commutation relations but introduces non-local interactions. In 3+1 dimensions, such techniques become increasingly cumbersome or inapplicable, making this strategy unattractive. Alternatively, other studies maintain Gauss’s law explicitly by introducing large penalty terms to suppress unphysical states, an effective approach when only low-energy dynamics are of interest~\cite{Hauke:2013jga,Halimeh:2019svu,VanDamme:2020rur}.

The Gauss’s law plays a unique role in gauge theories, particularly in the initial-value problem - if the initial state satisfies Gauss’s law, it remains satisfied throughout the evolution, provided the dynamics preserve gauge invariance. As the generator of gauge transformations, Gauss’s law is therefore automatically enforced when the implementation is exactly gauge invariant. This reverses the usual logic - rather than imposing Gauss’s law to ensure gauge invariance, one can construct a fully gauge-invariant formulation in which the constraint is inherently satisfied. In this sense, gauge theories are not truly constrained systems. Moreover, Gauss’s law can itself serve as a mechanism for quantum error correction~\cite{Stryker:2018efp,Rajput:2021trn}.

On the spatial lattice, gauge links are essential to maintain gauge invariance in the presence of charged fields. In the continuum limit, the lattice commutation relation ($[E,U]=U$) is equivalent to the canonical commutation relation [E,A]=i, but their finite-dimensional approximations differ because they preserve different properties. A direct truncation maintains the commutation relation exactly but breaks unitarity, whereas no finite-dimensional representation can exactly preserve the canonical commutation relation. Both can be unified under the finite Weyl commutation relation, where the configuration and conjugate field bases are connected by a Fourier transform. This connection enables efficient implementation of both momentum and interaction terms, and allows the representation dimension to increase systematically while maintaining exact unitarity of time evolution.

Alternative formulations, such as Quantum Link Models~\cite{Chandrasekharan:1996ih,Brower:1997ha,Brower:2003vy,Banerjee:2012xg,Kasper:2015cca,Felser:2019xyv,Magnifico:2020bqt,Halimeh:2021ufh,Wiese:2021djl}, represent the commutation relations using finite-dimensional spin operators, often through the rishon decomposition of gauge fields into gauge singlets. Related constructions appear in the Loop–String–Hadron framework~\cite{Raychowdhury:2019iki,Davoudi:2020yln,Kadam:2022ipf}, based on the Schwinger boson (prepotential) formalism~\cite{Mathur:2004kr}, though it applies only to non-Abelian theories.

Among these approaches, the unitary representation of the gauge link is the most practical for extending to large representation dimensions and lattice volumes, essential for approaching the continuum limit. Studies in lower dimensions~\cite{Kasper:2015cca,Zache:2021ggw,Maiti:2023kpn} support this scalability. Large representation dimensions are also crucial for real-time simulations, as they correspond to higher accessible energy scales necessary to capture physical processes.

The study focuses on the real-time evolution.
The initial state preparation is another big challenge, usually specified for different problems.
Here we only test the quantum circuit with simple initial states, and leave the general initial state for future works.
But the quantum error mitigation techniques are crucial at the present stage, so we present a comparison among different methods.
The paper is organised as follows:
Section \ref{sec:qed} and Section \ref{sec:rep} introduce the digitisation of the QED Hamiltonian on the lattice and in the Hilbert space respectively.
Section \ref{sec:qubit} is the heart of the paper; this is where we put the QED in qubit. 
We present the result in Section \ref{sec:res} and summarise in Section \ref{sec:sum}.

\section{Hamiltonian Approach of QED }
\label{sec:qed}

The natural quantisation scheme for the quantum computation is the canonical quantisation on the Hamilton, which, for QED, has long been done and is well presented in the textbook.
Here for practical computation, we need to put both the continuum space-time and the infinite Hilbert space into finite dimensional representation.

\subsection{Hamiltonian and Constraint}

The gauge theory is special in the canonical quantisation because its Hamilton system is constrained. 
The naive canonical quantisation can not be directly assigned to the fields, as it will be in conflict with the constraints.
Instead, the right commutation relations are obtained via the gauge fixing and  the Dirac bracket \cite{Weinberg:1995mt}.
As there are many ways to introduce the gauge fixing, QED can be studied under many different conditions, which ultimately are supposed to deliver the same physical results, i.e., those of gauge invariant observables.
The most intrinsic one for the Hamiltonian approach is the temporal gauge fixing, where we have the Hamilton 
\begin{align}
H\!=\!\!\!\int\!\! \ud^3x\! \left[
\frac{1}{2}\bm{\Pi}^2 \!+ \!\frac{1}{2}\bm{B}^2 
\!+\!\bar\psi\gamma^i\left(\partial_i-igA_i\right)\psi\!+\!m\bar\psi\psi
\right]
,
\end{align}
with the canonical commutation relations
\begin{align}
\left[\!A_i(x),\Pi^j(y)\!\right]\!\!=\!i\delta^j_i\delta^3\!(x\!-\!y)
,
\left\{\!\psi(x),\psi^\dagger(y)\!\right\}\!\!=\!\delta^3\!(x\!-\!y)
,
\end{align}
where the electric field $\bm{E}=-\bm{\Pi}$ and the magnetic field $\bm{B}=\nabla\times \bm{A}$.
The temporal gauge fixing does not fully resolve the constraints, and 
the remaining constraint is the Gauss's law,
\begin{align}
\nabla\cdot \bm{\Pi} +g\psi^\dagger \psi=0
.
\end{align}

On the other hand, the gauge theory is also very special when treated as a constrained theory.
Namely, the Gauss's law is a first class constraint, so related to the gauge transformation, and if the gauge invariance is maintained during the evolution, then the result will satisfy the constraint, the Gauss's law, automatically.
This is true both for the classical and quantum fields.
We will use this property to construct the qubit approach for QED.

There are also other popular gauge fixing conditions, such as
Coulomb gauge fixing, where only physical degrees, the transverse modes, are present,
and Axial gauge fixing, where the Gauss's law is solved out along one spatial direction.
We can in principle construct the qubit approach for any specific gauge fixing condition.
Generally for those approaches, the price of removing the gauge redundancy is the non-local interactions, which will require more gate operations in the circuit, as we can imagine.
This reflects the trade-off between the qubit and the gate-operation that we will encounter frequently.

It is also easy to show from the temporal gauge fixing that we can obtain other gauge fixing conditions via some unitary transformations \cite{Lenz:1994tb}.
This is expected as these different descriptions are in theory equivalent, although their implementations are far more different.
In comparison to other gauge fixings, there is no non-local interaction involved in the temporal gauge fixing.
The presence of the constraint seems its drawback, but the Gauss's law constraint is not really a constraint once the gauge transformation is preserved.
Another advantage for the temporal gauge is that it can be easily adopted in the non-Abelian theory and on the lattice.

\subsection{Lattice discretisation}

The continuous space can firstly be put into the spatial lattice, on which the lattice field theory is well defined.
As it is well known from Wilson \cite{Wilson:1974sk}, the gauge link is the only way to preserve the gauge invariance on the lattice with the presence of charged field.
The Hamilton on the lattice is
\begin{align}
&H= \ud^3x \sum_x\Bigg[
\frac{1}{2}\sum_i\Pi_i^2 \!+\! \frac{1}{4g^2}\sum_{i,j}\frac{2\!-\!P_{ij}\!-\!P_{ji}}{(\ud x^i\ud x^j)^2}
\\ &
\!+\!\sum_i\bar\psi(x)\gamma^i\frac{U_i(x)\psi(x\!+\!i)\!-\!U_i^\dagger(x\!-\!i)\psi(x\!-\!i)}{2\ud x}\!+\!m\bar\psi\psi
\Bigg]
\nonumber
.
\end{align}
Here no sum over repeated index, unless the sum symbol is present.
The spatial plaquette is constructed from four gauge links,
$P_{ij}(x)= U_i(x)U_j(x+i)U_i^\dagger(x+j)U_j^\dagger(x)$.
The continuum limit can be obtained if we treat the gauge link 
$U_i(x)\sim \exp\left(-igA_i(x)\ud x\right)$
and take $\ud x\to 0$.
The non-trivial commutation relations are then in the discrete form
\begin{gather}
[\Pi_i(x),U_j(y)]=-\frac{g}{\ud^2 x}\delta_{ij}\delta_{x,y}U_j(y)
,\\
\{\psi_\alpha(x),\psi^\dagger_\beta(y)\}=\frac{1}{\ud^3x}\delta_{\alpha\beta}\delta_{x,y}
.
\end{gather}
The Gauss's law on a single site is
\begin{align}
G(x)\equiv \sum_i \frac{\Pi_i(x)-\Pi_i(x-i)}{\ud x} + g\psi^\dagger(x)\psi(x) =0
,
\end{align}
which commutes with every single term in the Hamiltonian.
Therefore, the Gauss's law will not be broken when we split the exponential of the Hamilton into a product of exponentials of each elementary term in the Lie-Trotter-Suzuki expansion, as long as each elementary term is {\it exactly} represented.

Due to the fermion doubling problem, we can either consider Wilson fermion or the staggered fermion \cite{Kogut:1974ag}.
The staggered fermion allocates the spinor degrees on neighboring spatial sites, and we can recover the Dirac fermion when we take neighboring spatial sites into one.
For simplicity, we are going to use the staggered fermion for demonstration, and it is straightforward to extend the analysis to the Wilson fermion.
For more details on the lattice, see Appendix  \ref{app:lattice}, where we also give a complete discussion on the 3D cube and 2D square that we are going to work on.

\section{Digitisation of the representation}
\label{sec:rep}

All non-trivial commutation relations happen to the fields on the same spatial site.
After rescaling, such as the one in Appendix \ref{app:lattice}, we deal with the closed commutator algebra
\begin{align}
[\Pi,~U]=-U,\quad
[\Pi,~U^\dagger]=U^\dagger,\quad
[U,~U^\dagger]=0
.
\end{align}
The electric field operator is Hermitian, while the gauge-link field operator is unitary.
The first two of them indicate the gauge-link field is a ladder operator in the electric basis, i.e., the eigen-basis of the electric field,
\begin{align}
\Pi\!=\!\left(
\begin{array}{ccccc}
\resizebox*{10pt}{!}{\rotatebox[origin=c]{-45}{$\cdots$}} &   \\
  &$-$1 &  &  \\
  &  & 0 &  \\
  &  &  & 1 \\
  &  &  &  & \resizebox*{10pt}{!}{\rotatebox[origin=c]{-45}{$\cdots$}} \\
\end{array}
\right)
\!,~
U\!=\!\left(
\begin{array}{ccccc}
 \resizebox*{10pt}{!}{\rotatebox[origin=c]{-45}{$\cdots$}}\!\!\! &  \resizebox*{10pt}{!}{\rotatebox[origin=c]{-45}{$\cdots$}}  \\
  &\0 & 1  &  \\
  &  & \0 & 1  \\
  &  &  &\0  &\resizebox*{10pt}{!}{\rotatebox[origin=c]{-45}{$\cdots$}} \\
  &  &  &  &\resizebox*{10pt}{!}{\rotatebox[origin=c]{-45}{$\cdots$}}\\
\end{array}
\right)
\!.
\end{align}
The eigenstates and the matrices above are infinite dimensional in both plus and minus directions, as the third commutation relation does not provide a stop on either direction.
In practice, we have to adopt a truncation and this will introduce violations on the commutation relations.
Depending on where the violation occurs, there are two different choices.

If we apply a hard truncation, the unitarity of the gauge-link field will be violated, 
\begin{align}
[\Pi,~U']=-U',\quad
[\Pi,~U'^\dagger]=U'^\dagger,\quad
[U',~U'^\dagger]\approx 0
, \\ \nonumber
{\rm e.g.,~}
\Pi=\left(
\begin{array}{cccc}
$-$2 \\
  &$-$1 &  &  \\
  &  & 0 &  \\
  &  &  & 1 \\
\end{array}\right)
,\quad
U'=\left(
\begin{array}{cccc}
 \0 & 1 & \0 & \0 \\
 \0 & \0 & 1 & \0 \\
 \0 & \0 & \0 & 1 \\
 \0 & \0 & \0 & \0 \\
\end{array}\right)
,
\end{align}
in $4\times 4$ matrices.
As it shows, the truncated ladder matrix is not a unitary matrix.
On the other hand, it is easy to restore the unitarity, but the price to pay is  that the commutation relation between the electric field and the gauge-link field will be violated, 
\begin{align}
[\Pi,~U]\approx -U,\quad
[\Pi,~U^\dagger]\approx U^\dagger,\quad
[U,~U^\dagger]= 0
, \\ \nonumber {\rm e.g.,~}
\Pi=\left(
\begin{array}{cccc}
$-$2 \\
  &$-$1 &  &  \\
  &  & 0 &  \\
  &  &  & 1 \\
\end{array}\right)
,\quad
U=\left(
\begin{array}{cccc}
 \0 & 1 & \0 & \0 \\
 \0 & \0 & 1 & \0 \\
 \0 & \0 & \0 & 1 \\
  1 & \0 & \0 & \0 \\
\end{array}\right)
.
\end{align}
To be precise, the commutation relation between $\Pi$ and $U$ still hold true to some extent, when we rephrase the unbounded operators into bounded ones,
\begin{align}
[\Pi,U]\approx - U~  &\Leftrightarrow  e^{-i\alpha \Pi} U e^{i\alpha \Pi} = e^{-i\alpha} U
\end{align} 
with $\alpha=2\pi k/N$ for $k \in {\mathbb Z}$, where $N$ is the dimension of the matrix.
Since the electric field operator is the generator of the gauge transformation, this means the gauge transformation is restricted to $N$ discrete phases, thus  the $U(1)$ gauge theory is reduced to $ Z_N$ theory in the unitary matrix approach.
Only in $N\to \infty$ limit,  the $U(1)$ gauge theory will be restored \cite{Horn:1979fy}.

The unitarity will eventually pay off when we put matrices into gate-operations.
We note that in order to keep the gauge invariance, we have to implement every elementary term exactly.
For a matrix $M$ of a large dimension, the precise qubit implementation of the rotation will reply on the eigen-decomposition, 
\begin{align}
e^{-i\theta f(M,M^\dagger)}={\mathscr U}^\dagger e^{-i\theta f(\Lambda,\Lambda^\dagger)}{\mathscr U}
,\quad
M={\mathscr U}^\dagger \Lambda {\mathscr U}
.
\end{align}
Since Pauli decomposition of the diagonal matrix $\Lambda$ consists of Pauli strings of tensor products of $I$ and $Z$ only and all of them commute, the rotation term in the middle can be done exactly.
So, the challenges are: (1) whether we can find the eigen-decomposition and (2) whether we can implement the unitary ${\mathscr U}$ exactly.
Fortunately, both challenges can be effectively resolved using the unitary representation of the gauge link field.

\noindent
{\bf Unitary Matrix and Fourier Transform:}
It is useful to notice that the unitary representation of the gauge-link field coincides with both the standard digitisation of the scalar field \cite{Jordan:2012xnu} and the finite representation of Weyl commutation relation \cite{Weyl, Santhanam}.
For the bosonic field, there is no reliable finite dimensional representation for the unbounded operators in the Canonical Commutation Relation.
As adopted in \cite{Jordan:2012xnu}, we can represent $A$ and $\Pi$ in their relevant eigen-bases, which are discrete and finite dimensional.
The two bases are then connected via the discrete Fourier Transform, which is reminiscent of the Fourier Transform in the continuum theory.
In this way, the commutation relation is violated, but the Hermiticity of $A$ and $\Pi$ are conserved, and most importantly we have the Fourier Transform that can be implemented to any large representation dimension.
If we take the exponential of $A$, as to get the gauge link $U=e^{-iA}$, we obtain the unitary approach in the gauge-link commutation relation.
Unlike in the gauge-link where the truncated ladder matrix can maintain exactly the commutation relation, there exist no finite dimensional matrices that replicate the Canonical Commutation Relation.

\begin{center}
\includegraphics[width=\linewidth]{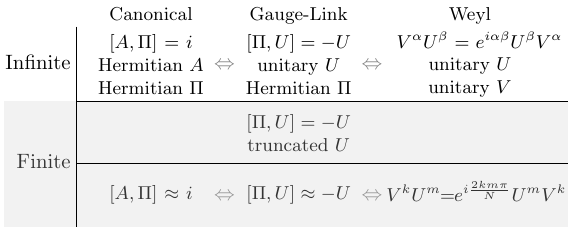}
\end{center}

If we further take the exponential of $\Pi$, i.e., $V= \exp\left(-i\Pi\right)$, we reach the Weyl formulation of the canonical commutation relation, $V^\alpha U^\beta=e^{i\alpha \beta}U^\beta V^\alpha$ for real numbers $\alpha,~\beta$.
A finite dimensional representation exists via Sylvester's generalization of Pauli matrices with the shift matrix and the clock matrix.
For the gauge theory, we can choose
\begin{gather}
U=- F^\dagger \exp\left(-i\frac{2\pi}{N}\Pi\right) F
,\quad
V= \exp\left(-i\frac{2\pi}{N}\Pi\right)
,\nonumber\\
V^kU^m=U^mV^ke^{i2mk\pi/N}.
\end{gather}
where $\Pi$ is a diagonal integer matrix.
For example, when $N=4$, we get
\begin{align}
\Pi=\left(
\begin{array}{rrcc}
$-$2 & \0 & \0 & \0 \\
 \0 &$-$1 & \0 & \0 \\
 \0 & \0 & 0 & \0 \\
 \0 & \0 & \0 & 1 \\
\end{array}
\right)
,\quad
U=\left(
\begin{array}{cccc}
 \0 & 1 & \0 & \0 \\
 \0 & \0 & 1 & \0 \\
 \0 & \0 & \0 & 1 \\
 1 & \0 & \0 & \0 \\
\end{array}
\right)
,\\
F=\frac{1}{\sqrt{4}}\left(
\begin{array}{cccc}
1 & 1& 1 &1 \\
1 & \omega & \omega^2 &\omega^3 \\
1 & \omega^2 & \omega^4 &\omega^6 \\
1 & \omega^3 & \omega^6 &\omega^9 \\
\end{array}
\right)
,\quad
\omega=e^{i2\pi/4}
.
\end{align}

Although the three representations are equivalent in the continuum limit, they can have different individual discretisation schemes, which may or may not be equivalent.
Here the choice of unitary $U$ is the same approach that all of the three can share.
Unlike in the canonical or gauge-link commutation relation, the equality in the Weyl formulation is still kept in the finite dimensional representation.

Given the even dimension $N=2^n$, we may introduce a constant $1/2$ to the electric field $\Pi$, so that the spectrum is more symmetric between positive and negative values, as in \cite{Klco:2018zqz}.
In this case, $V^N=-I$ and we will deal with $Z_N$ with a twisted boundary condition.
Also a fermionic-like spectrum may lead to some non-trivial difference \cite{Halimeh:2021ufh}.
So we will not do the symmetrization here.

The advantage of following the Weyl commutation relation is that the setup can well be constructed up to the $N\to \infty$ limit, under which we will not only recover the Canonical Commutation Relation \cite{Weyl} but also restore the $U(1)$ gauge \cite{Horn:1979fy}.
On the other hand, the Fourier transform matrix $F$ which connects $U$ and $V$ here allows us to switch the eigen-basis between the electric and gauge potential fields.
For future use, we will call the bases the electric basis and the potential basis respectively.
Notice the potential basis will also be the eigen-basis of the magnetic field, but generally it is not yet gauge invariant and should be distinguished with the magnetic basis in the literature, such as \cite{Bauer:2021gek, Burbano:2024uvn, Clemente:2022cka, Kane:2022ejm, Bender:2020ztu, Haase:2020kaj}.
This easy switch between the electric basis and the potential basis turns out to be crucial when we construct the quantum circuit.
Most importantly, depending on the quantum algorithm of Quantum Fourier Transform, we can carry out the implementation to any large N.

\noindent
{\bf Truncated Ladder Matrix:}
In the result of the hard truncation, the gauge link field is represented by the direct truncated ladder matrix $U'$, which is defective, and thus does not admit the eigen-decomposition directly.
In the end, it is the exponent $f(M,M^\dagger)$ that we wish to decompose.
For the exponent in some specific shape, we can achieve the eigen-decomposition of the exponent via the Singular Value Decomposition of the matrix $M=\mathscr{V}S\mathscr{W}^\dagger$, with ${\mathscr V}$, ${\mathscr W}$ unitary matrices and $S$ the diagonal matrix with non-negative real numbers \cite{Davoudi:2022xmb}.
This includes the gauge-fermion interaction which will be in a form
\begin{align}
\!\left(\! \begin{array}{cc}  0 & M \\ M^\dagger\! & 0 \end{array} \!\right)\!  =
\!\left(\! \begin{array}{cc} \mathscr{V} & 0 \\ 0 & \mathscr{W} \end{array} \!\right)\!
{\mathbf H}
\!\left(\! \begin{array}{cc}   S &  \\  & $-$S \end{array} \!\right)\!
{\mathbf H}^\dagger\!
\!\left(\! \begin{array}{cc} \mathscr{V}^\dagger\! & 0 \\ 0 & \mathscr{W}^\dagger\! \end{array} \!\right)\!
,
\label{eq:svd1}
\end{align}
where ${\mathbf H}$ is the Hadamard operation.
The Singular Value Decomposition of course also applies to the unitary matrix.
For instance, for $U=UII$, we have $\mathscr{V}=U$ and $\mathscr{W}=S=I$.
We will use the formula to perform the rotation of the gauge-fermion interaction.
The truncated ladder matrix can then be done, depending on the unitary matrix,  i.e., $U'=U\Lambda I$, with $\Lambda=diag(0,1,1,\cdots)$, thus $\mathscr{V}=U$, $\mathscr{W}=I$ and $S=\Lambda$.

The plaquette term can also be done based on this, if we split a ladder matrix into two nilpotent matrices, given that for the nilpotent matrix $N^2=0$,
\begin{align}
\!\left(\!\begin{array}{cc} &  \!N\!+\!N^\dagger\! \\  N\!+\!N^\dagger\! & \end{array}\!\right)\! 
\!=\!
\!\left(\!\begin{array}{cc} &  \!\!N \\  N^\dagger\! & \end{array}\!\right)\!
\!\left(\!\begin{array}{cc} &  \!\!N^\dagger\! \\  N & \end{array}\!\right)\!
.
\end{align}
Therefore, we can introduce an ancilla to perform the plaquette interaction.
We give a detailed explanation in Appendix \ref{app:rep}.

In the non-Abelian theory, we can also have the directly truncated matrix representation of the gauge link field, whose commutation relations with electric fields are exact.
In comparison, the unitary representation of the gauge link is no longer available in the non-Abelian theory, neither will be the Quantum Fourier Transform.
See Appendix \ref{app:ext} for more information.

\section{QED in Qubit}
\label{sec:qubit}

\begin{figure*}[t!]
\includegraphics[width=\linewidth]{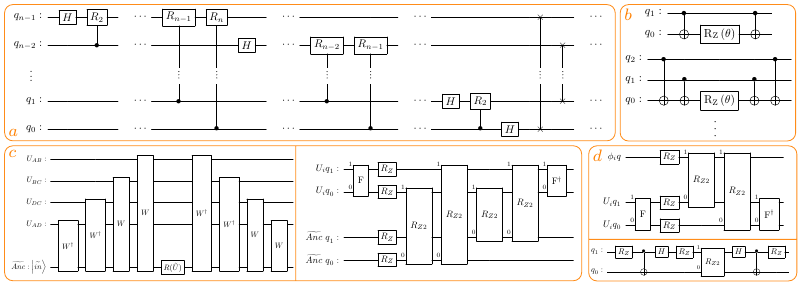}
\caption{%
An illustration of the quantum circuits that are used in the method.
(a) Quantum Fourier Transform.
(b) multi-Z rotations.
(c) The plaquette rotation in the ``stator'' method: (L) the four links and the ancilla; (R) the W-operation illustrated with 2-qubit for the gauge degree.
(d) The gauge-fermion rotation: (Upper) the unitary transformation which decouples the gauge degree; (Lower) the two-fermion rotation.
}
\label{fig:circuits}
\end{figure*}

Based on the setup listed in the previous section, it is straightforward to put the matrix representation into qubit.

\subsection{Qubit for the gauge field}

Via the Fourier transform, we can simply switch between the electric basis and the potential basis.
In the eigen-basis, the relevant operator is in a diagonal shape, and a diagonal matrix allows a Pauli $(I,Z)$ decomposition, and can easily and precisely be implemented in qubit.
In practice, we have the electric field
\begin{align}
\Pi \!=\! -\frac{1}{2} \sum_{j=0}^{n-1} 2^j Z_j - \frac{1}{2}
,~ {\rm with}~
Z_j \!\equiv \!
\underset{\mathclap{\substack{\uparrow \\ \scalebox{1}{\tiny $(\!n\!\!-\!\!1\!)$}{\text -}{\rm th}}}}{I}
 \otimes \cdots 
\overset{\mathclap{\substack{\tiny j{\text -}{\rm th~ from~ right} \\ \downarrow}}}{Z}
\cdots \otimes
\underset{\mathclap{\substack{\uparrow \\ \tiny 0{\text -}{\rm th}}}}{I}
\end{align}
where the shorthand $Z_j$ stands for only the $j{\text -}$th term is $Z$, not $I$, and the unitary matrix $U$ which is diagonalized via the Fourier transform
\begin{align}
U &= F^\dagger \exp\left(-i\frac{2\pi}{N}\left(\Pi+\frac{N}{2}\right)\right) F
.
\end{align}
$N\equiv 2^n$.
As a circulant matrix, the unitary matrix $U$ can be put into quantum circuit directly \cite{Pueschel:1998zzo}, but its eigen-decomposition representation is more useful to switch between the two bases.
In the potential basis, the $U$-relevant terms is in a diagonal shape, and the relevant Trotterization consists of only rotations of $Z$'s, so can be implemented precisely.
This precise implementation is crucial, as the Gauss's law is satisfied only when each elementary Trotter term is done exactly.

Thus, to implement the evolution in high dimensional representation space, we rely on two basic operations.
The first is the Quantum Fourier Transform, which is a fundamental quantum algorithm and can be called directly on different platforms.
The second is the multi-$Z$ rotation, in a form of 
$\exp\left(-i\frac{\theta}{2} Z_{m}\otimes Z_{m-1}\otimes \cdots \otimes Z_0\right)$.
With Controlled-NOT (or CX) gate, the rotation can easily be constructed from the single qubit rotation recursively, for example, see Fig. \ref{fig:circuits}(b).
There are also other different arrangements, but generally the number of CX-gates is fixed, as $2m$ CX-gates for the $(m+1)$-Z rotation.
For an eigenstate, the multi-Z rotation does not alter the state, but merely introduces an overall phase term according to the parity.
This is why the CX-gates are always in pair to compensate any flip.
With an ancilla, we can simply register the parity, but without reinstating the ancilla after the rotation.
Thus in this way we may reduce the number of CX-gates
\cite{Hastings:2014wyq,Farrell:2022wyt}.

\noindent
{\bf Plaquette:}
The plaquette interaction can straightforwardly be implemented in the potential basis, where the gauge link field is diagonal.
The rotation can be done exactly via the multi-Z rotations.
The total number of the required $CX$-gates will amount to
\begin{align}
\sum_{k=0}^{4(n-1)} \binom{4(n-1)}{k} 2(k+3)=
\frac{N^4}{8}\left(1+2\log_2N\right)
.
\end{align}
There also exists in the literature an excellent method, called ``stator" \cite{Reznik_2002,Zohar:2016iic,Zohar:2016wmo,Bender:2018rdp}, which can reduce the number of CX gates with the use of ancillary qubits.
For matrix $C$ satisfying $C^N=I$,
we can construct an entanglement between the target qubits and the ancillary qubits via
\begin{gather}
W=\exp\left(i\frac{N}{2\pi} \ln(C)\otimes \ln(\tilde V)\right)
,\nonumber\\{\rm such~ that}\quad
W^\dagger \left(\tilde U\right)^m W = \left(C\otimes \tilde U\right)^m
.
\end{gather}
Then, by selecting for the ancillary qubits the eigenstate $|\widetilde{in}\rangle$, which satisfies $ \tilde U |\widetilde{in}\rangle= |\widetilde{in}\rangle$ and can be created through the Quantum Fourier Transform,
we can obtain
\begin{align}
W^\dagger f\left(\tilde U\right) W |\Psi\rangle \otimes |\widetilde{in}\rangle 
= f\left(U\right) |\Psi\rangle \otimes |\widetilde{in}\rangle
\end{align}
for any analytic function $f(x)$.
This suggests we can achieve an rotation on the target qubits but without performing the rotation on the target qubits directly.
Moreover, it is easy to check with two entanglements in sequence
\begin{align}
W_2^\dagger W_1^\dagger \left(\tilde U\right)^m W_1 W_2 = \left(C_2\otimes C_1\otimes \tilde U\right)^m
.
\end{align}
Therefore, the plaquette interaction can be attained with a single rotation on the ancilla.
A demonstration for the circuit implementation is shown in Fig. \ref{fig:circuits}(c).
As we can see, there are four pairs of $W$-operations in the plaquette and each takes $2\left(\log_2(N)\right)^2$ $CX$-gates.
Also, the $W$-operation does not take longer than two-$Z$ rotations.
In fact, the most resource consuming part is the rotation $e^{-i\frac{\theta}{2}(\tilde U+\tilde U^\dagger)}$, which takes 
$\frac{N}{2}\left(\log_2N-1\right)$ CX-gates.
Therefore, compared to the direct approach, the stator method will take extra $\log_2N$ qubits, but reduce the number of gates from $O(N^4)$ to $O(N)$.
For this reason, we are going to adopt the method in our approach.

\subsection{Qubit for fermion}

The fermion fields can be put into Pauli matrices via the Jordan–Wigner mapping,
for example,
\begin{align}
\phi_A=            \frac{X_A-iY_A}{2}
,\quad             
\phi_B= Z_A        \frac{X_B-iY_B}{2}
,\quad \cdots            
\end{align}
A typical gauge-fermion interaction term can be split into three two-field-operations through, e.g.,
\begin{align}
&\exp\left(
-i\frac{\theta}{2}
\left[i\phi_P^\dagger U_{PQ}\phi_Q-i\phi_Q^\dagger U^\dagger_{PQ}\phi_P\right]
\right)
\nonumber\\=&
R_{P}\exp\left(
-i\frac{\theta}{2}
\left[i\phi_P^\dagger \phi_Q-i\phi_Q^\dagger \phi_P\right]
\right)R_{P}^\dagger
,
\end{align}
with the unitary transformation as in Eq. \ref{eq:svd1}, 
\begin{align}
R_{P}
= \left(\begin{array}{cc} U_{PQ} &  \\  & I \end{array}\right)
= \exp\left(\phi_P^\dagger\phi_P \ln(U_{PQ})\right)
,
\end{align}
working on $(n+1)$ qubits.
The two-fermion rotation in the middle involves only two qubits, and can simply be performed via multi-Z rotations.
In practice, the $Z$-string might be longer, depending on the separation between $P$ and $Q$ in the Wigner-Jordan map.  
It is easy to extend the analysis to Dirac or Wilson fermions, and the gauge-fermion interaction can always be performed via the procedure above.

\subsection{Initial states and Quantum Error Mitigation (QEM) methods}

In this work, we focus on the real-time evolution of QED, specifically to address the challenge of finding a scalable expansion of the representation dimension.
The preparation for a general initial state is another challenge, if not a bigger one,  which is especially the case with the constraint of the Gauss's law.
Fortunately, it is enough to use simple initial state to test out the implementation of the evolution.
We are going to use eigenstates in the electric basis for the initialization, and postpone the study of the general initial state preparation for future work.

In theory, initial physical modes will be kept to be physical during the evolution.
By physical modes, we mean the modes with eigenvalue $0$ for {\it all} Gauss's law operators on spatial sites.
Besides the physical modes, allowed are also those of eigenvalue $\pm N$ or $\pm 2N$ for the Gauss's law operators in the $Z_N$ gauge theory.
We will use modulo modes to refer those extra modes.
With the presence of the quantum error, we will also encounter the error modes which appear merely due to the quantum error.
Given the quantum error is still the significant factor in the nowadays quantum computer, we ought to find ways to extract the signal from the noise.
We consider two general ways to do so.

\noindent
{\bf Post-selection method:}
Since we know the error modes are completely due to the presence of the quantum error, we can naively reduce the noise by restricting to use only those allowed modes. 
One challenge of the approach comes from the curse of dimensionality.
On a single site, the allowed modes, including both the physical and modulo modes, take only $1/N$ of all modes, as shown in Appendix \ref{app:phy}.
Considering the whole lattice, there will be a further suppression to the power of the total number of spatial sites, and the {\it exact} allowed modes take a tiny fraction of the whole Hilbert space.
As a result, when a quantum error happens, the probability that the allowed modes are changed into error modes is exponentially high.
We need either an exponentially large sample size or an exponentially small quantum error rate to counter the effect when we apply the post-selection to the final measurement.
One way out is to relax the constraints. 
We may use the Gauss's law to classify the states in the final measurement and then to set a threshold to include in the calculation also those states with smaller violation of the Gauss's law.
On the other hand, we may give up using all the constraints simultaneously, but instead only apply locally related constraints for the observable.
However, from practical tests, we find the post-selection methods, even with some relaxation, do not work efficiently, especially in a large system.

\noindent
{\bf Calibration method:}
Following \cite{Farrell:2023fgd}, we are to adopt the so-called Operator Decoherence Renormalization method to analyze the final measurement. 
There are quantum error mitigation techniques integrated in the IBM quantum computers, specially Dynamical decoupling and Pauli twirling, which can be enabled directly via qiskit command lines.
After the Pauli twirling, the quantum error can be described by a Pauli channel \cite{Dur:2005obk}, where the measured value of an operator can be related to its exact value via
\begin{align}
O_{\rm mea}  = \sum_{i}^{4^n} \eta_i {\rm Tr}\left[ O P_i\rho   P_i\right]
 =c_{O} O_{\rm exa} 
.
\end{align}
Here the $ P$'s are tensor products of Pauli operators ($ I$, $ X$, $ Y$ or $ Z$) acting on $n$-qubit, and the observable $ O$, which is also a tensor product of Pauli operators,  commutes or anti-commutes with $ P$'s, as there are only two options for the Pauli operators.
Notice the constant $c_O$ merely relies on the operator, not the state.
For this work, we select $ O=Z_i$ for every qubit and calibrate the coefficients $c_O$ individually, from which the electric fields and fermion charges can then be constructed accordingly.
In practice, for each run we are interested in, there is one more accompanying run to calibrate the result, where two of them share the same circuit, but different parameters.
In order to compute the coefficients, the exact result of the run for the calibration shall be known in advance, and we find $\ud t=0$ is a good option, as the final result should be the same as the initial state if without error.
Technically, the twin runs share the same circuit with merely different rotation angles.
Since the rotations act only on a few single-qubit gates and the quantum noise is dominated by the two-qubit gates, the two runs are expected to have the same error channel \cite{Farrell:2023fgd}, therefore, the same coefficient $\eta_i$, which can simply be extracted from the trivial run of $\ud t=0$.
For finite $\ud t$, we obtain the mean and the standard error
\begin{gather}
 O_{\ud t} = O_{{\rm mea},\ud t}\frac{ O_{{\rm exa},0}}{ O_{{\rm mea},0}}
,\\
\sigma_{\ud t} =| O_{\ud t}|\sqrt{
\left(\frac{\sigma_{{\rm mea},\ud t}}{ O_{{\rm mea},\ud t}}\right)^2
+\left(\frac{\sigma_{{\rm mea},0}}{ O_{{\rm mea},0}}\right)^2
}
.
\label{eq:prop}
\end{gather}
Due to the error propagation formula above, there may be a large error accumulated from the denominators when the measured values are close to zero. 
For example, for $O=Z$ the calibration method may not work properly when the measurement is distributed evenly on $Z=\pm 1$ states.
This is especially the case when the quantum error rate is high.
In the worst case, the final distribution is evenly distributed without regard of the operations beforehand.
Therefore, the calibration method works when the quantum error is not wildly large.

\section{Simulations and results}
\label{sec:res}

We are to perform the code for the real-time evolution of QED on the cube (3D), for which we will first provide a detailed explanation with one plaquette (2D) where all types of interactions are present.
The resource usage for the quantum computation of one Trotter step can be listed as follows
\begin{center}
\includegraphics[width=0.8\linewidth]{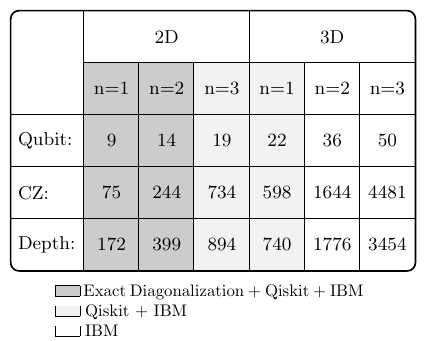}
\end{center}
where Qubit, CZ and Depth are individually the total number of qubits, the total number of controlled-Z gates and the depth of the circuit that the QED code generates for the situations.
The number of CZ gates and the circuit depth here should be taken in a loose sense as the numbers are obtained from a particular IBM quantum computer, ${\it ibm\_aachen}$, and may vary on different machines and optimizations.
The $n$ is the number of qubits for the gauge field representation, and in the study for both the plaquette (2D) and the cube (3D) configurations we choose $n=1,2,3$, so each gauge degree in $N=2,4,8$ dimensional representation.
The Exact Diagonalization is available when the qubit count is 9 or 14, .
When the count goes up to 19 or 22, we can not perform the Exact Diagonalization due to the memory shortage on the desktop, but we can still do the Trotter evolution on the wavefunctional.
When the number is 36 or 50, even the Qiskit simulation is no longer available because of the exponential growth of the memory requirement,  and we can only rely on the quantum computer.

\begin{figure*}[t!]
\begin{center}
\includegraphics[width=\linewidth]{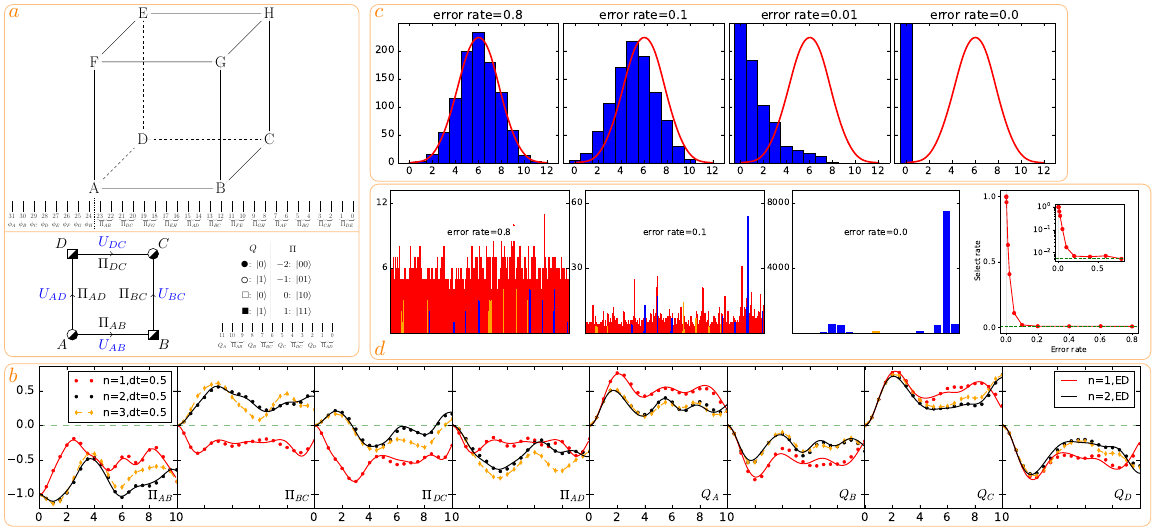}
\caption{%
(a) The illustration of (Upper) the 3D cube and (Lower) the 2D square.
The qubit arrangements are also displayed separately, illustrated with 2-qubit for each gauge degree.
An explanation of the fermion occupation and the electric basis is also present.
(b) The observables, including electric fields and fermion charges, evolve along time on the 2D square.
The solid lines stand for the results from Exact Diagonalization, while the dots refer to Qiskit simulations of $\ud t=0.5$.
(c) The distance (or the bit-flip number) plot with varying quantum error rate in the trivial runs, simulated via Qiskit.
(d) The histogram of the final states in the target runs, simulated via Qiskit.
The right-most plot is the select rate versus error rate.
}
\label{fig:sim}
\end{center}
\end{figure*}

\subsection{The plaquette}

For the plaquette, we select an allocation of qubits which depicts most of the geometry of the square, as shown in Fig. \ref{fig:sim}(a).
The electric field strength $\Pi$ and the fermion charge $Q$ are denoted into qubit as shown above, where to demonstrate we have used 2-qubit for each gauge degree.
Therefore, an eigenstate can be encoded into a binary string of 12 bits, to which its decimal can also be used as shorthand, for example,
\begin{align}
(2738) &\equiv | 101010110010 \rangle,
\quad
\Pi_{AB}=-1,
\nonumber\\
(3250) &\equiv | 110010110010 \rangle,
\quad
\Pi_{AB}=0,
\\\nonumber
(3762) &\equiv | 111010110010 \rangle,
\quad
\Pi_{AB}=1,
\end{align}
with all other $\Pi$'s and $Q$'s vanishing.
The only difference among the three states above is $\Pi_{AB}$.
Usually, it is the second state $(3250)$ that is considered as the physical state with vanishing open boundary condition. 
In fact, all three states can be regarded as physical states in the presence of different external electric fields, such as
\begin{center}
\includegraphics[width=\linewidth]{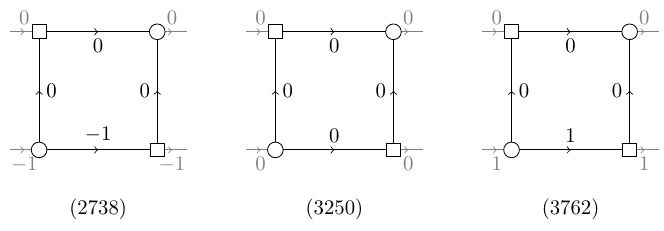} 
\end{center}
Since the definition of physical states varies with the background, a versatile code should not exclude any state.
Physically, if we put the three states above as initial states individually and evolve them with the same Trotterization procedure, they will evolve in separate parts of the phase space, and never cross each other.
One advantage of not integrating out the Gauss's law is that for different constant background, we only need to change the initial state according to the Gauss's law, but not the Hamiltonian.

In this Quantum Mechanical example, the Schwinger effect, the pair production in the presence of the electric field, can be observed.
In fact, the Schwinger effect can already be captured from 1+1D, and a spatial square of staggered fermion is not the best place to study the Schwinger effect with uniform external electric field, due to the geometric allocation of the staggered fermion, in comparison to the Wilson fermion.
But the staggered fermion is perfect to study the non-uniform electric field, where we also expect the magnetic field in the situation, which starts to appear only from the 2D configuration,  the plaquette $P\sim e^{-ig B\ud^2x}$.
There are two ways to define the magnetic field, 
\begin{align}
B=-\frac{{\rm Im}(\langle P\rangle )}{g\ud^2 x}
,\quad {\rm or}\quad
B=i\frac{\langle \ln(P)\rangle }{g\ud^2 x}
,
\end{align}
either from the imaginary or from the logarithmic part of the plaquette, as both functions are well defined in the unitary representation of the gauge link.
By definition, we only need to measure $Z$-strings in the potential basis.
For a single eigenstate in the electric basis that we have used for the initialization, this will always be zero, as after the Fourier transform all eigenstates are  equally available in the potential basis.
Therefore, we will postpone the study of the magnetic field to future more sophisticated initial state.

\noindent
{\bf Qiskit simulations:}
We first demonstrate with ED and Qiskit simulation what are the changes in the evolution of the observables with increasing qubit number $n$, for which we choose the electric fields $\Pi_{AB}$, $\Pi_{BC}$, $\Pi_{DC}$, $\Pi_{AD}$ and the fermion charges $Q_A$, $Q_B$, $Q_C$, $Q_D$, as shown in Fig. \ref{fig:sim}(b).
To compare, we choose a common simple initial state that all $n=1,2,3$ can have, that is, $\Pi_{AB}=-1$ and all others zero.
Firstly, as we see in the $n=1$ case $\Pi_{AB}$ increases immediately from $-1$ to $0$, as there are only two eigenstates, $\Pi=-1$ and $0$, present in the case.
A better resolution is achieved when we increase $n$.
In fact, we can spot a dip in $\Pi_{AB}$ shortly after the initialization in both $n=2,3$ that $n=1$ could not capture.
That limited energy range in $n=1$ also has an effect in $\Pi_{BC}$ and $\Pi_{DC}$, which can only decrease in the very beginning, although the right evolution is just the opposite.
Except those obvious inconsistencies, different $n$'s match each other in all observables, especially between $n=2$ and $3$.
Moreover, with the larger $n$, the discrepancy happens more later, and we expect the convergence to QED in the limit $n ~{\rm or}~N\to\infty$.
A study of the convergence with the large spin representation of the Quantum Link Models in 1+1D can be found in \cite{Kasper:2015cca,Zache:2021ggw,Maiti:2023kpn}.
In the plot, the dots stand for the measurements from the Trotter evolution of the wavefunctional.
For $n=1,2$, we can also perform Exact Diagonalization.
For a comparison, we have used $\ud t=0.5$ for the Trotter method, whose results are already very close to the ones in Exact Diagonalization.
We are going to stick to $\ud t=0.5$ in following simulations, as it also provides a longer time step.
For a more detailed study on $\ud t$, see Appendix \ref{app:dt}.

We can also perform Qiskit simulation with the quantum error included, and this can be achieved in AerSimulator where the quantum error rate can be varied via the parameter of the depolarizing error probability of CX-gate.

First of all, it is simple and interesting to study in the trivial runs the error rate from the number of bit flips, also known as Hamming distance.
The distance between two eigenstates can be easily computed as the number of different bits between the two states when expressed in binary strings.
We apply the distance measurement to $\ud t=0$, where without error the final state should be exactly the initial state.
For example, for a 12-qubit initial state 
$|111010110010 \rangle$,
there is 1 flip in $|\gray{0}11010110010 \rangle$ and 4 in $|\gray{0001}10110010 \rangle$.
If there is no error, the final distribution is a delta function on 0.
With larger and larger error, the distribution will be smeared to long distance range.
The worst scenario is the final state is totally random.
A clear comparison with different quantum error rates is present in Fig. \ref{fig:sim}(c), where as we see with increasing error rate, the distribution becomes closer to the random case, whose envelope is plotted in red to guide the eyes.

Next, we provide the distribution of all states in the final measurement after one Trotter step ($\ud t=0.5$) in the presence of different error rates.
The initial state is chosen to be $(3762)$, which is still the dominating state in the final measurement, as shown in the no-error one in Fig. \ref{fig:sim}(d).
The relevant physical states are indicated in blue, the modulo states in orange and the error states in red.
On the right most plot is a check on the post-selection, where the error rate is the parameter in Qiskit and the select rate is the ratio of the count of allowed modes over the total number of shots.
The green dashed line in the right-hand side plot is the ratio of all allowed modes over all possible modes.
As we see the post-selection rate is very small when the quantum error rate is large.
This will become even worse when the system size becomes larger, as the ratio of the exact allowed modes over all modes is exponentially suppressed by the spatial volume. 
Therefore, with a decreasing error rate, the select rate will go up in a very steep slop, which can already be spotted in the plot.

\begin{figure*}[t!]
\begin{center}
\includegraphics[width=\linewidth]{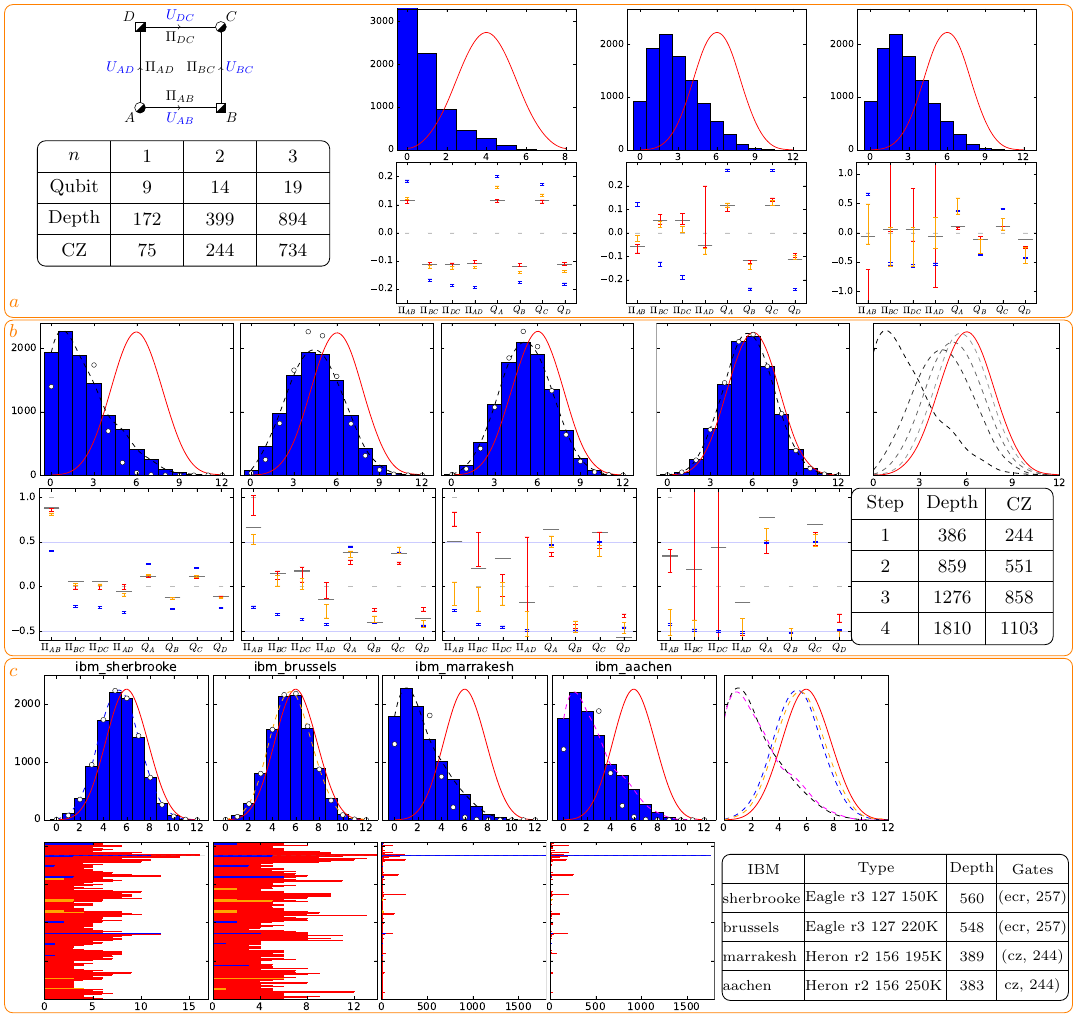}
\caption{
(a) Results for $n=1,2,3$ (Left, Middle, Right) on IBM quantum computers: the distance (Upper) and the observables (Lower), where the blue, red and orange error bars denote respectively the results from the direct, calibrated and post-selected measurements, and for comparison, the exact values are drawn as the gray bars.
(b) Results for 1, 2, 3, and 4 Trotter steps, of $n=2$. 
(c) Results from different platforms, with $n=2$ and one Trotter step.
(Upper) the distance plot;
(Lower) the histogram of the final measurements, with the count on x-axis and the state on y-axis. 
}
\label{fig:IBM2d}
\end{center}
\end{figure*}

\noindent
{\bf On IBM:}
We have carried out one-Totter-step evolution for the 2D plaquette with $n=1,2,3$ on IBM quantum computers, with the resource usage listed in Fig. \ref{fig:IBM2d}(a).
The number of circuit depth is obtained by setting optimization\_level=2 and may vary from platform to platform.
Both Dynamical Decoupling and Pauli Twirling are turned on during the simulations.
The final measurements are performed on all qubits except ancillary ones.

In Fig. \ref{fig:IBM2d}(a) plotted are the observables (with the initial values subtracted for plot), where the initial state has all $\Pi$'s and $Q$'s vanishing, except $\Pi_{AB}=-1$.
As we can see, all the direct measurements (in blue error bars) have big deviations from the exact values (in gray bars).
But the deviations have almost been alleviated under the calibration (in red error bars).
The improvement is more obvious in $n=1$ and $2$ cases, where the quantum error is small with shallower circuit.
In $n=3$, the large quantum error can easily be spotted both from the distance plot and from the big red error bars in the measurement.
The results from the post-selection are also presented in orange error bars.
The post-selection results posted here are of the full constraints.
We have also investigated the post-selection method by relaxing the constraints, but none of them work better than the one with full constraints in this small system.

\noindent
{\bf Multi Trotter steps:}
We have also tested for $n=2$ up to four Trotter steps, with a resource usage and final result shown in Fig. \ref{fig:IBM2d}(b).  
The direct measurements of the observables are shown in blue error bars and the calibrated in red.
The exact values along time are plotted in the gray lines.
As we see, the quantum error becomes larger and larger with the increasing Trotter steps.
In the distance plots, the distributions of the final measurements become closer and closer to the random case with more Trotter steps.
Meanwhile, the error bars after the calibration also become bigger.
Notice in the four-Trotter-step case where the distribution is close to the random one, the measurements are close to $\pm 0.5$.
Given the expressions of $\Pi$'s and $Q$'s on Pauli $Z$'s, this indicates the expectation values of all $Z$'s are around zero, which is another sign of the random distribution and the reason of the big error from the propagation formula Eq. (\ref{eq:prop}).

\noindent
{\bf From IBM Eagle to IBM Heron:}
We have carried out a comparison of the same code for $n=2$ one-Trotter-step evolution among different IBM platforms.
A summary of resource usage and result is present in Fig. \ref{fig:IBM2d}(c).
There is a big difference between IBM Eagle and Heron platforms, that is, the native two-qubit gate operation, ecr vs cz.
The circuit depth is largely decreased from the upgrade of the quantum processor, as well as the running time. 
There is a noticeable improvement from Eagle r3 to Heron r2, which can be seen from both the distance plot (upper panel) and the histogram of the states (lower panel).

\subsection{The cube}

\begin{figure*}[t!]
\includegraphics[width=\linewidth]{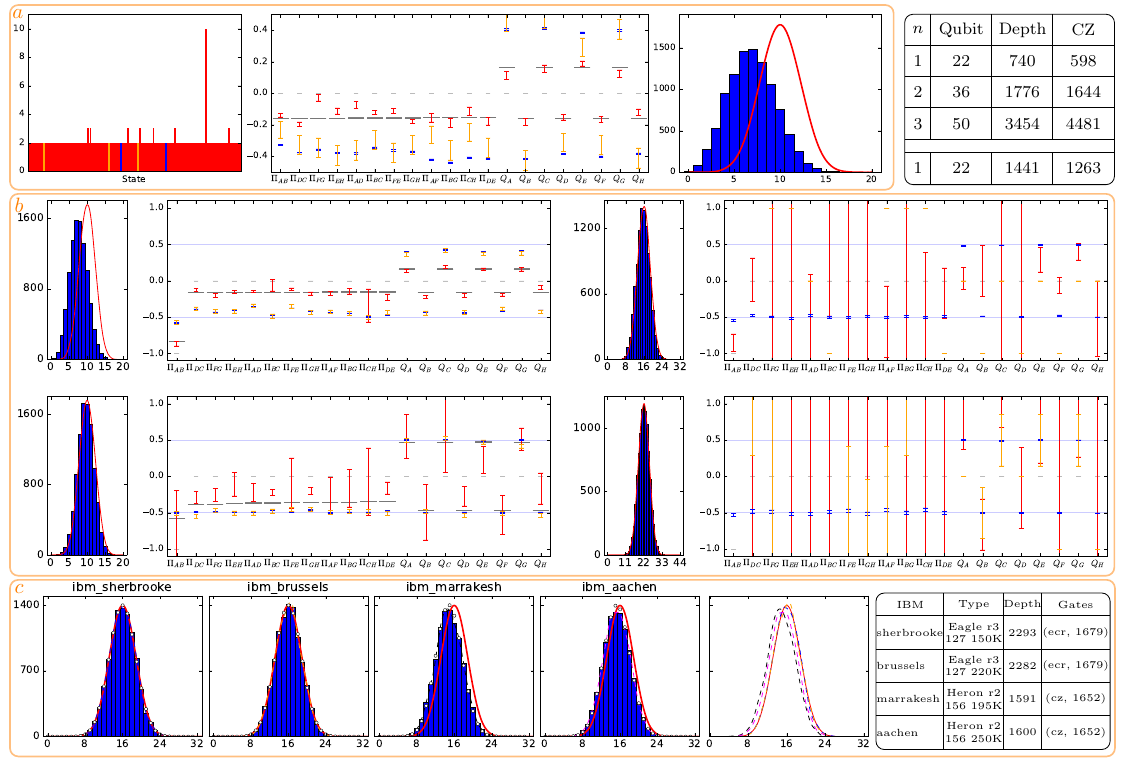}
\caption{
On 3D cube: (a) Results for $n=1$  on IBM quantum computers: the histogram of the final states (Left), the distance (Right) and the observables (Middle), where the blue, red and orange error bars denote respectively the results from the direct, calibrated and post-selected measurements, and for comparison, the exact values are drawn as the gray bars.
(b) Simulations with initial state of $\Pi_{AB}=-1$.
(Left) $n=1$ with one- (Upper) and two- (Lower) Trotter steps;
(Right) $n=2$ (Upper) and $n=3$ (Lower) with one Trotter step.
(c) The distance plot of results from different platforms, with $n=2$ and one Trotter step.
}
\label{fig:IBM3d}
\end{figure*}

The smallest layout of the QED on the 3D lattice is a cube with the open boundary condition, which consists of 8 spatial sites.
There are 8 fermion fields, each of which requires 1-qubit for the representation.
There are 12 gauge fields.
If, for example, we choose 2-qubit to represent a single gauge field, the setup will then consume 32 qubits.
We also need ancilla qubits to perform the plaquette interaction.
Notice the two opposite plaquettes on the cube can be operated simultaneously if they call on different ancilla qubits, as there will be no overlap between the two.
So in practice we may need two sets of ancilla qubits to enable parallel operations of the plaquette interaction, each of which requires the same amount of qubits as the gauge field.
On the whole, the total number will then go up to 36 qubits.
The qubit number is already too large for the exact state vector method.
Even with the matrix product state method, the Qiskit will require about 75 TB of memory on the classical computer to simulate the general Trotter steps.
The quantum computer is almost the only option for the circuit of the cube.
In comparison, if we reduce 2-qubit for each gauge field to 1-qubit, the total number will be 22, and in this case, the Qiskit simulation is available on the desktop.
We have carried out the one-Totter-step for the cube on IBM quantum computers, with the resource usage listed in the right upper corner in Fig. \ref{fig:IBM3d}.

\vskip 2pt
\noindent
{\bf n=1:}
Together with extra 2 ancillary qubits, the circuit can still be simulated with the classical computer, and the exact values can then be compared with the results from the quantum computer where the quantum error kicks in.
An example is shown in Fig. \ref{fig:IBM3d}(a).
The histogram of the states in the final measurement is displayed in the left-hand side plot.
There are 10,000 shots in use, and to plot concisely, we only plot the states with at least 2 shots collected.
The final measurement looks almost full of errors.
But in fact, the distribution is much better than the worst scenario, which can be seen from the distance plot on the right-hand side.
The observables are shown in the middle, where again the blue error bars refer to the direct measurements and the calibrated results are shown in red error bars.
The exact values are shown in gray lines, and in addition, the results from the post-selection are also presented in orange error bars.
The post-selection results posted here are of the full constraints.
We have also investigated the post-selection method by relaxing the constraints, but no post-selection method works better than the calibration method in this comparatively large system.

For $n=1$ with 22 qubits, we have also carried out one more Trotter step.
A comparison between the results of one and two Trotter steps is shown in left-hand side of Fig. \ref{fig:IBM3d}(b), for which the initial state is chosen with $\Pi_{AB}=-1$.
With more than thousand CZ-gate operations, the quantum error accumulated in the two-Trotter-step evolution is already big enough, shown as the large error bars in the calibration method.
The big noise can also been seen from both the distance plot where the distribution is close to the random one and the measurement where the values are close to $\pm 0.5$ that cause big error bars.

\vskip 2pt
\noindent
{\bf n=2,3:}
It is straightforward to increase the qubit number to represent the gauge degree, as done in the 2D case.
The big difference for 3D is that now the qubit number becomes so big that the simulation on the classical computer is no long available, and we can only carry out the simulation on the quantum computer.
The results of one-Trotter-step evolution for $n=2,3$ on the IBM quantum computer are shown in the right-hand side of Fig. \ref{fig:IBM3d}(b).
First notice the huge error bars in the calibrated results, compared to the $n=1$ case.
We can even spot the results become much worse from $n=2$ to $3$.
Also in the distance plots, the distributions are very close to the random ones, which indicate big quantum error accumulated.
Meanwhile in the measurement, the values of the observables are too close to $\pm 0.5$, which is the direct reason for the huge errors in the calibrated results.
The quantum error accumulated on the quantum computer is too big to be reliable.

\vskip 2pt
\noindent
{\bf From IBM Eagle to IBM Heron:}
For 3D $n=2$, we have done a similar comparison on different platforms, with the result shown in Fig. \ref{fig:IBM3d}(c).
Due to the large circuit depth, the accumulated quantum error is so big that the final distribution is almost uniform among possible states, and we cannot distinguish the signal from the noise.
In the distance plot, the distributions are very close to the random one.
In fact, what we have attained on Heron is not the worst error scenario, as we can see an improvement with the upgrade of the quantum platform.
We can extract a signal with a large error bar on IBM Heron platforms, whose distribution of distance is very similar to what we have seen in the result of 2D case on Eagle platforms.
Therefore, we might anticipate on next-generation quantum platforms we might be able to get reliable result for the 3D case, as we have achieved in 2D case from the platform upgrade.

\section{Conclusions}
\label{sec:sum}

We have demonstrated that QED can be implemented into qubit, with the quantum algorithm similar to the scalar field theory \cite{Jordan:2012xnu}.
For the extra constraint, namely the Gauss's law, accompanied with the gauge theory, we follow  the fact that the Gauss's law will be dissolved automatically when the implementation is completely gauge invariant.
To maintain the gauge transformation on the spatial lattice, we have to adopt the gauge link field, as the only way to preserve the gauge symmetry exactly.
Among the too many finite approximations to the commutation relation, we stick to a special one, which is consistent with both the scalar field setup \cite{Jordan:2012xnu} and the finite representation of the Weyl commutation relation \cite{Weyl}.
The advantage of the approach is that the setup can be easily extended to a large representation dimension $N$ of the gauge field.
The large $N$ is crucial for the setup, as we can not only cover a large energy range for the simulation but also recover the compact QED from the $Z_N$ gauge theory in the limit $N\to \infty$.

We regard this setup as a standard approach, as it is a direct application of the quantum algorithm of the scalar field theory \cite{Jordan:2012xnu}.
In fact, other approaches can also reply on the unitary approach.
For example, we can also implement the truncated ladder matrix, based on the unitary matrix of the gauge link.
But in comparison, there exist too many extra operations in Singular Value Decomposition, for which reason we do not follow in this study.

We have also studied the different Quantum Error Mitigation methods with the setup.
The states that satisfy all Gauss's law constraints are extremely rare, exponentially suppressed by the spatial volume, as the Gauss's law happens on each spatial site.
Therefore, a post selection with the exact allowed modes is highly disfavoured.
The calibration method provides a way out.
Due to the error, the states will be spreading around the allowed modes, and the calibration method will make good use the spreading distribution.

We have tested out the codes on IBM quantum computers. 
We can obtain a clear signal on Heron platform for the 2+1D setup with a few Trotter steps.
For the 3+1D setup, we can see an obvious improvement from Eagle to Heron platforms.
For the large system, we can only reply on the quantum computer, due to the exponential growth in the memory demand for the classical computer.
We are now at a tantalizing point where the upgrade of the quantum processor in the near future might be able to give a reliable result of the large simulation performed exclusively on quantum computer.

\section*{Acknowledgement}

We would like to thank Debasish Banerjee, Antonio Barbalace, Andreas Juettner, Graham Van Goffrier, Paul Kelly, and Ali Rezeai for insightful discussions. Research of BC at the University of Southampton
has been supported by the following research grants - STFC (Grant no. ST/X000583/1), STFC (Grant no. ST/W006251/1), and EPSRC (Grant no. EP/W032635/1), and research of ZM at the University of Southampton has been supported by EPSRC (Grant no. EP/W032635/1). We would like to acknowledge the quantum computational hardware resources provided by National Quantum Computing Center (NQCC) and IBM under NQCC's Quantum Computing Access Programme (QCAP).

\appendix

\section{More on Lattice}
\label{app:lattice}
The Gauss's law operator commutes with every single term in the Hamiltonian
\begin{gather}
[\Pi_i(x),G(y)]=0
,\quad
[P_{ij}(x),G(y)]=0
,\\
[\psi_\alpha^\dagger(x)U_i(x)\psi_\beta(x+i),G(y)]=0
,\\
[\psi_\alpha^\dagger(x)\psi_\beta(x),G(y)]=0
\end{gather}
Therefore, the Gauss's law will not be broken during the following Trotterization
\begin{align}
&H= \ud^3 x\sum_x\left[ {\mathcal H}_{\Pi}(x) +{\mathcal H}_{P}(x) +{\mathcal H}_{I}(x) +{\mathcal H}_{Q}(x) \right]
,\\
&e^{-i\ud t H}\approx \prod e^{-i\ud t\ud^3x {\mathcal H}_\Pi} e^{-i\ud t\ud^3x {\mathcal H}_P} e^{-i\ud t\ud^3x {\mathcal H}_{I}} e^{-i\ud t\ud^3x {\mathcal H}_{Q}}\nonumber
\end{align}
as long as each elementary term is wholly represented.

Due to the fermion doubling problem, we can either consider Wilson fermion, 
which adds the Wilson mass term to the Hamiltonian, 
\begin{align}
H_{W}= \ud^3x\sum_x \Bigg[ 
-\frac{r_W\ud x}{2}\sum_i\bar\psi(x)\\
\frac{U_i(x)\psi(x+i)-2\psi(x)+U_i^\dagger(x-i)\psi(x-i)}{\ud^2 x}
\Bigg]
\end{align}
or the staggered fermion, given by Kogut and Susskind \cite{Kogut:1974ag}
\begin{gather}
H_{KS}=\ud^3x\sum_x\Bigg[
\frac{1}{2}\Pi^2
+\frac{2-P-P^\dagger}{4g^2(\ud x)^4}
\nonumber\\
+i\sum_j\eta^j(x)\frac{\phi^\dagger(x)U_j(x)\phi(x+j)-\phi^\dagger(x+j)U^\dagger_j(x)\phi(x)}{2\ud x_j}
\nonumber\\
+m(-)^n\phi^\dagger(x)\phi(x)
\Bigg]
\end{gather}
where $\eta^1=1,~\eta^2=(-)^{n_1},~\eta^3=(-)^{n_1+n_2}$ with $x_i=n_i\ud x$.
On each spatial site, there is only one fermion field for the staggered fermion, and in comparison, there are four for the Wilson fermion.
The staggered fermion allocates the spinor degrees on neighboring spatial sites, and we can recover the Dirac fermion when we take neighboring spatial sites into one.
For simplicity, we are going to use the staggered fermion for demonstration, and it is straightforward to extend the analysis to the Wilson fermion.

\subsubsection{The 3D cube} 

The smallest 3-dimensional (3D) spatial setup is a cube with open boundary condition, see Fig. \ref{fig:sim}(a) for an illustration.
By rescaling to dimensionless variables via
$\frac{(\ud x)^2 }{g}\Pi \to \Pi
,\quad
(\ud x)^{3/2} \phi \to \phi$,
we eventually deal with
\begin{gather}
H=\frac{1}{\ud x}\Bigg[
\frac{\alpha}{2}\big(
 \Pi_{AB}^2
+\Pi_{DC}^2
+\Pi_{FG}^2
+\Pi_{EH}^2
+\Pi_{AD}^2
+\Pi_{BC}^2
\nonumber \\ 
+\Pi_{FE}^2
+\Pi_{GH}^2
+\Pi_{AF}^2
+\Pi_{BG}^2
+\Pi_{CH}^2
+\Pi_{DE}^2
\big)
\nonumber \\ 
-\frac{1}{4\alpha}\big(
 P_{ABCD}
+P_{FGHE}
+P_{ADEF}
+P_{BCHG}
\nonumber \\ 
+P_{ABGF}
+P_{DCHE}
+h.c.
\big)
\nonumber \\ 
+\frac{i\phi_A^\dagger U_{AB}\phi_B+h.c.}{2}
+\frac{i\phi_D^\dagger U_{DC}\phi_C+h.c.}{2}
\nonumber \\ 
+\frac{i\phi_E^\dagger U_{EH}\phi_H+h.c.}{2}
+\frac{i\phi_F^\dagger U_{FG}\phi_G+h.c.}{2}
\nonumber \\ 
+\frac{i\phi_A^\dagger U_{AD}\phi_D+h.c.}{2}
-\frac{i\phi_B^\dagger U_{BC}\phi_C+h.c.}{2}
\nonumber \\ 
+\frac{i\phi_F^\dagger U_{FE}\phi_E+h.c.}{2}
-\frac{i\phi_G^\dagger U_{GH}\phi_H+h.c.}{2}
\nonumber \\ 
+\frac{i\phi_A^\dagger U_{AF}\phi_F+h.c.}{2}
-\frac{i\phi_B^\dagger U_{BG}\phi_G+h.c.}{2}
\nonumber \\ 
+\frac{i\phi_C^\dagger U_{CH}\phi_H+h.c.}{2}
-\frac{i\phi_D^\dagger U_{DE}\phi_E+h.c.}{2}
\nonumber \\ 
+\beta\phi_A^\dagger\phi_A
-\beta\phi_B^\dagger\phi_B
+\beta\phi_C^\dagger\phi_C
-\beta\phi_D^\dagger\phi_D
\nonumber \\ 
+\beta\phi_E^\dagger\phi_E
-\beta\phi_F^\dagger\phi_F
+\beta\phi_G^\dagger\phi_G
-\beta\phi_H^\dagger\phi_H
\Bigg]
\end{gather}
with dimensionless parameters
$\alpha = g^2$ and
$\beta  = m\ud x$.
Without causing any confusion, we have utilized double subscripts, e.g., $U_{AB}$, to denote the gauge link and four subscripts to refer the plaquette, e.g., $P_{ABCD}\equiv U_{AB}U_{BC}U_{DC}^\dagger U_{AD}^\dagger$.
The canonical commutation relations become
\begin{gather}
[\Pi_i,U_j]=-\delta_{ij}U_j
,\quad
[\Pi_i,U^\dagger_j]=\delta_{ij}U^\dagger_j
,\\
\{\phi_i,\phi^\dagger_j\}=\delta_{ij}
\end{gather}
The Gauss's law can be expressed as
\begin{align}
G_A&=~~ \Pi_{AB}+\Pi_{AD}+\Pi_{AF}+\phi^\dagger_A\phi_A ~~~~~~=0
,\nonumber\\
G_B&=  -\Pi_{AB}+\Pi_{BC}+\Pi_{BG}+\phi^\dagger_B\phi_B-1=0
,\nonumber\\
G_C&= -\Pi_{DC}-\Pi_{BC}+\Pi_{CH}+\phi^\dagger_C\phi_C ~~~~~=0
,\nonumber\\
G_D&=~~\Pi_{DC}-\Pi_{AD}+\Pi_{DE}+\phi^\dagger_D\phi_D-1=0
,\\
G_E&=~~\Pi_{EH}-\Pi_{FE}-\Pi_{DE}+\phi^\dagger_E\phi_E ~~~~~~=0
,\nonumber\\
G_F&=~~\Pi_{FG}+\Pi_{FE}-\Pi_{AF}+\phi^\dagger_F\phi_F-1 ~=0
,\nonumber\\
G_G&= -\Pi_{FG}+\Pi_{GH}-\Pi_{BG}+\phi^\dagger_G\phi_G ~~~~~=0
,\nonumber\\
G_H&= -\Pi_{EH}-\Pi_{GH}-\Pi_{CH}+\phi^\dagger_H\phi_H-1=0
.\nonumber
\end{align}
The Trotterization can be implemented as
\begin{align}
e^{-i\ud t H} \approx&
\exp\left( -i\frac{\theta_Q}{2}\left[ \phi_A^\dagger\phi_A  \right] \right)
\cdots
\label{eq:q}
\\&
\exp\left( -i\frac{\theta_\Pi}{2}\left[ \Pi_A^2 \right] \right)
\cdots
\label{eq:e}
\\&
\exp\left( -i\frac{\theta_I}{2}\left[ i\phi_A^\dagger U_{AB}\phi_B+h.c \right] \right)
\cdots
\label{eq:i1}
\\&
\exp\left( -i\frac{\theta_P}{2}\left[ P_{ABCD} +h.c \right] \right)
\cdots
,
\label{eq:p}
\end{align}
with the angles
\begin{align}
\theta_P=-\frac{\ud t}{2\alpha\ud x}
,~
\theta_I=\frac{\ud t}{\ud x}
,~
\theta_\Pi=\frac{\alpha\ud t}{\ud x}
,~
\theta_Q=\frac{2\beta\ud t}{\ud x}
,
\end{align}
where the angles might differ by a sign on different fermion sites.

\subsubsection{The 2D plaquette}

In fact, all types of interactions above have already been included in a single plaquette.
It is also easier to demonstrate the treatment with the 2D plaquette.
We may start from theory of two spatial dimensions and obtain
\begin{gather}
H=\frac{1}{\ud x}
\Bigg[
\frac{\alpha}{2}\left(\Pi_{AB}^2+\Pi_{BC}^2+\Pi_{DC}^2+\Pi_{AD}^2\right)
\nonumber \\ 
+\frac{2-P_{ABCD}-P_{ABCD}^\dagger}{4\alpha}
\nonumber \\ 
+i\frac{\phi_{A}^\dagger U_{AB}\phi_{B}-\phi_{B}^\dagger U^\dagger_{AB}\phi_{A}}{2}
-i\frac{\phi_{B}^\dagger U_{BC}\phi_{C}-\phi_{C}^\dagger U^\dagger_{BC}\phi_{B}}{2}
\nonumber \\ 
+i\frac{\phi_{D}^\dagger U_{DC}\phi_{C}-\phi_{C}^\dagger U^\dagger_{DC}\phi_{D}}{2}
+i\frac{\phi_{A}^\dagger U_{AD}\phi_{D}-\phi_{D}^\dagger U^\dagger_{AD}\phi_{A}}{2}
\nonumber \\ 
+\beta\phi_{A}^\dagger\phi_{A}
-\beta\phi_{B}^\dagger\phi_{B}
+\beta\phi_{C}^\dagger\phi_{C}
-\beta\phi_{D}^\dagger\phi_{D}
\Bigg]
\end{gather}
where the rescaling varies in 2D as $\frac{(\ud x )}{g}\Pi \to \Pi ,~~ (\ud x) \phi \to \phi$, and $\alpha = g^2\ud x ,~ \beta = m\ud x$.
As we can see, it has the similar appearance as the 3D case in $\alpha$, $\beta$ parameters, although the rescaling is different.

The single plaquette is in practice established in the following way:
\begin{center}
\includegraphics[width=0.45\linewidth]{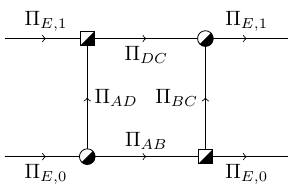} 
\end{center}
For the staggered fermions, the charge is defined separately for the particle and anti-particle
\begin{align}
\scalebox{1.75}{$\circ$}/\scalebox{1.75}{$\bullet$}
,\quad
Q=\frac{Z+1}{2}
=\Bigg\{\begin{array}{l}
\scalebox{1.75}{$\bullet$},\quad 1,\quad |0\rangle = \Big(\begin{array}{c} 1 \\ 0\end{array}\Big)
\\
\scalebox{1.75}{$\circ$},\quad 0,\quad |1\rangle = \Big(\begin{array}{c} 0 \\ 1\end{array}\Big)
\end{array}
,\\
\square/\blacksquare
,\quad
Q=\frac{Z-1}{2}
=\Bigg\{\begin{array}{l}
\square,\quad ~~0,\quad |0\rangle = \Big(\begin{array}{c} 1 \\ 0\end{array}\Big)
\\
\blacksquare,\quad -1,\quad |1\rangle = \Big(\begin{array}{c} 0 \\ 1\end{array}\Big)
\end{array}
\end{align}
The Gauss's law is expressed as
\begin{align}
G_{A}&= ~~\Pi_{AB}+\Pi_{AD}+\phi^\dagger_{A}\phi_{A} -\Pi_{E,0}=0
,\\
G_{B}&= ~~ \Pi_{BC}-\Pi_{AB}+\phi^\dagger_{B}\phi_{B}-1+\Pi_{E,0}=0
,\\                              
G_{C}&= -\Pi_{BC}-\Pi_{DC}+\phi^\dagger_{C}\phi_{C}+\Pi_{E,1}=0
,\\                              
G_{D}&= ~~ \Pi_{DC}-\Pi_{AD}+\phi^\dagger_{D}\phi_{D}-1-\Pi_{E,1}=0
\end{align}
where we have introduced the external/background electric fields.
It is interesting to observe that the external fields appear only in the Gauss's law, but do not alter the interactions.
This means for different background setups we only need to choose the initial state to meet the Gauss's law, but not change the following Trotter evolution process.

\section{More on the physical modes}
\label{app:phy}

In the section, we are going to compute the number of physical states under the Gauss's law constraint.
Given the electric field operator $\Pi$ is diagonal in the electric basis, e.g., $\Pi\in [-N/2,N/2-1]$, 
the electric part of the Gauss's law on a spatial site admits
\begin{align}
{\nabla}\cdot {\Pi} \in [-3N+3,3N-3]
.
\end{align}
For a specific eigenvalue $k=\nabla \cdot \Pi$, the total number of possible combinations is
\begin{align}
f(k)\!=\!\!\!\!\!\sum_{i=0}^{\left[\frac{k+3N-3}{N}\right]}\!\!\!\! (-)^i\binom{6}{i} \binom{k+3N-3+6-1-iN}{6-1}
,
\end{align}
where the bracket means the floor function which returns the largest integer less than or equal to $(k+3N-3)/N$. 
The distribution is centered at zero, with the maximum value
\begin{align}
f(0)=\frac{11N^5+5N^3+4N}{20}
,
\end{align}
and can be approximated with a normal distribution
\begin{align}
f(k)\approx \frac{A}{\sqrt{2\pi}\sigma}\exp\left(-\frac{k^2}{2\sigma^2}\right)
,
\end{align}
provided with the parameters
\begin{align*}
A\!\approx\! N^6,~ \frac{A}{\sqrt{2\pi}\sigma} \!\approx\! f(0) 
\to \sigma \!\approx\!  \frac{20 N^6}{\sqrt{2\pi}(11N^5+5N^3+4N)}
.
\end{align*}
When $N\to \infty$, the distribution is also known as the Irwin-Hall distribution, specifically in this case as the sum distribution of six uniform distributions.

The number of $f(0 {\rm ~mod~} N)$ can be obtained by summing over $k=0,~\pm N,~\pm 2N$, 
\begin{align}
f(0 {\rm ~mod~} N)=N^5.
\end{align}
In fact, we will get the same number for any $f(k {\rm ~mod~} N)$.
Therefore, the number of the allowed modes, i.e., the modulo modes, does not alter when the fermion is present to shift $k$.
The allowed modes take $1/N$ of all modes, and among the allowed modes, the physical modes take about $11/20$ accordingly to the central value. 
This happens to a single spatial site.
On the whole lattice, we expect the ratio becomes further suppressed to the power of the total number of spatial sites.

\section{Digitization for Truncated Ladder Operators}
\label{app:rep}

\noindent
{\bf Singular Value Decomposition:}
In the result of the hard truncation, the matrix $U'$ is defective, and thus does not admit the eigen-decomposition directly.
There exist other ways where we can get the diagonalization for the rotation.
Given the Singular Value Decomposition (SVD) of the matrix
$M=\mathscr{V}S\mathscr{W}^\dagger$
where ${\mathscr V}$, ${\mathscr W}$ are unitary matrices and $S$ is the diagonal matrix with non-negative real numbers, we can diagonalize the following matrix
\cite{Davoudi:2022xmb}
\begin{gather}
\left( \begin{array}{cc}  0 & M \\ M^\dagger & 0 \end{array} \right)
=
\left( \begin{array}{cc} \mathscr{V} & 0 \\ 0 & \mathscr{W} \end{array} \right)
\left( \begin{array}{cc}  0 & S \\ S & 0 \end{array} \right)
\left( \begin{array}{cc} \mathscr{V}^\dagger & 0 \\ 0 & \mathscr{W}^\dagger \end{array} \right)
=\nonumber \\
\left( \begin{array}{cc} \mathscr{V} & 0 \\ 0 & \mathscr{W} \end{array} \right)
{\mathbf H}
\left( \begin{array}{cc}   S &  \\  & -S \end{array} \right)
{\mathbf H}^\dagger
\left( \begin{array}{cc} \mathscr{V}^\dagger & 0 \\ 0 & \mathscr{W}^\dagger \end{array} \right)
,
\end{gather}
where ${\mathbf H}$ is the Hadamard operation 
\begin{align}
{\mathbf H}\equiv
\frac{1}{\sqrt{2}}\left( \begin{array}{cr} 1 & 1 \\ 1 & -1 \end{array} \right)
,\quad
{\mathbf H} X {\mathbf H} = Z
.
\end{align}
This can be used to diagonalize the rotation when the exponent is of specific shape.
In fact, that particular shape above is usually the result of the fermion field.
For example, the Jordan–Wigner mapping leads to 
\begin{align}
\phi \!=\! \frac{X-iY}{2} \!=\! \left( \begin{array}{cc}  \0 & \0 \\ 1 & \0 \end{array} \right)
,~
\phi^\dagger \!=\! \frac{X+iY}{2} \!=\! \left( \begin{array}{cc}  \0 & 1 \\ \0 & \0 \end{array} \right)
.
\end{align}

\noindent
{\bf Gauge-Fermion Interaction:}
Consider the rotation of the gauge-fermion interaction term
\begin{align}
R=\exp\left(-i\frac{\theta_I}{2} \left[ i\phi_A^\dagger U_{AB}\phi_B +h.c.  \right] \right)
.
\end{align}

When the gauge link $U_{AB}$ is represented by the unitary matrix $U$, the SVD is trivial
\begin{gather}
M=-iU_{AB}\otimes 
\frac{X_B-iY_B}{2}
=UI\otimes \frac{I-Z}{2}(-iX)
,\\
\mathscr{V}=U\otimes I
,\quad
\mathscr{W}=iI\otimes X
,\quad
S=I\otimes \frac{I-Z}{2}
,
\end{gather}
with which the gauge-fermion rotation can be implemented exactly
\begin{gather}
R
=
\left( \begin{array}{cc}  U &  \\  & I \end{array} \right)
\left( \begin{array}{cc}  I &  \\  & iX \end{array} \right)
{\mathbf H}
\exp\Bigg(-i\frac{\theta_I}{2}
Z\otimes I\otimes \frac{I-Z}{2}
\Bigg)\nonumber\\
{\mathbf H}^\dagger
\left( \begin{array}{cc}  I &  \\  & -iX \end{array} \right)
\left( \begin{array}{cc}  U^\dagger &  \\  & I \end{array} \right)
.
\end{gather}
In the unitary matrices above, $U$ is really $U\otimes I$ and $X$ should be $I\otimes X$.
Here we omit the identities in outer products hopefully without causing any confusion.

For truncated ladder operator, the SVD of the defective matrix can easily be  achieved with the unitary matrix, for example,
\begin{gather}
U'=\left(
\begin{array}{cccc}
 \0 & 1 & \0 & \0 \\
 \0 & \0 & 1 & \0 \\
 \0 & \0 & \0 & 1 \\
 \0 & \0 & \0 & \0 \\
\end{array}\right)
,\quad
U=\left(
\begin{array}{cccc}
 \0 & 1 & \0 & \0 \\
 \0 & \0 & 1 & \0 \\
 \0 & \0 & \0 & 1 \\
 1 & \0 & \0 & \0 \\
\end{array}\right)
,\nonumber\\
\Lambda=\left(
\begin{array}{cccc}
 \0 & \0 & \0 & \0 \\
 \0 &  1 & \0 & \0 \\
 \0 & \0 &  1 & \0 \\
 \0 & \0 & \0 &  1 \\
\end{array}\right)
,\quad
U'=U\Lambda 
.
\end{gather}
We only need to modify the diagonal matrix above to obtain the rotation in the representation
\begin{gather}
R=
\left( \begin{array}{cc}  U &  \\  & I \end{array} \right)
\left( \begin{array}{cc}  I &  \\  & iX \end{array} \right)
{\mathbf H}
\exp\Bigg(-i\frac{\theta_I}{2}
Z\otimes \Lambda \otimes \frac{I-Z}{2}
\Bigg)
\nonumber\\
{\mathbf H}^\dagger
\left( \begin{array}{cc}  I &  \\  & -iX \end{array} \right)
\left( \begin{array}{cc}  U^\dagger &  \\  & I \end{array} \right)
.
\end{gather}

\noindent
{\bf Nilpotent matrix:}
The plaquette interaction does not fit into directly the expression above.
Ref. \cite{Davoudi:2022xmb} suggests we may first split the ladder matrix into two nilpotent matrices, $U'=U_E+U_O$, with $U_E^2=U_E^2=0$.
Notice for the nilpotent matrix $N$, there exist projectors to have
\begin{align}
\!\left(\!
\begin{array}{cc}
N+N^\dagger &  \\  & 0
\end{array}\!\right)\!
\!=\!
\!\left(\! \begin{array}{cc} {\mathcal P}_0 &  {\mathcal P}_1 \\ {\mathcal P}_1 & {\mathcal P}_0 \end{array}\!\right)\!
\!\left(\!\begin{array}{cc}
 & N \\ N^\dagger
\end{array}\!\right)\!
\!\left(\! \begin{array}{cc} {\mathcal P}_0 &  {\mathcal P}_1 \\ {\mathcal P}_1 & {\mathcal P}_0 \end{array}\!\right)\!
.
\end{align}
For example,
\begin{align*}
{\mathcal P}_0\!=\!\!\left(\!
\begin{array}{cccc}
  1 & \0 & \0 & \0 \\
 \0 & \0 & \0 & \0 \\
 \0 & \0 &  1 & \0 \\
 \0 & \0 & \0 & \0 \\
\end{array}\!\right)\!
\!,
U_E\!=\!\!\left(\!
\begin{array}{cccc}
 \0 &  1 & \0 & \0 \\
 \0 & \0 & \0 & \0 \\
 \0 & \0 & \0 &  1 \\
 \0 & \0 & \0 & \0 \\
\end{array}\!\right)\!
\!,
{\mathcal P}_1\!=\!\!\left(\!
\begin{array}{cccc}
 \0 & \0 & \0 & \0 \\
 \0 &  1 & \0 & \0 \\
 \0 & \0 & \0 & \0 \\
 \0 & \0 & \0 &  1 \\
\end{array}\!\right)\!
;\\
{\mathcal P}_0\!=\!\!\left(\!
\begin{array}{cccc}
  1 & \0 & \0 & \0 \\
 \0 &  1 & \0 & \0 \\
 \0 & \0 & \0 & \0 \\
 \0 & \0 & \0 &  1 \\
\end{array}\!\right)\!
\!,
U_O\!=\!\!\left(\!
\begin{array}{cccc}
 \0 & \0 & \0 & \0 \\
 \0 & \0 &  1 & \0 \\
 \0 & \0 & \0 & \0 \\
 \0 & \0 & \0 & \0 \\
\end{array}\!\right)\!
\!,
{\mathcal P}_1\!=\!\!\left(\!
\begin{array}{cccc}
 \0 & \0 & \0 & \0 \\
 \0 & \0 & \0 & \0 \\
 \0 & \0 &  1 & \0 \\
 \0 & \0 & \0 & \0 \\
\end{array}\!\right)\!
.
\end{align*}
In practice, the qubit representation of the projector matrices may not be an easy task.
We can skip its usage with the identity for the nilpotent matrix $A$
\begin{gather}
\exp\left(-i\frac{\theta}{2} \left(\begin{array}{cc} &  A+A^\dagger \\  A+A^\dagger & \end{array}\right) \right)
=\nonumber\\
\exp\left(-i\frac{\theta}{2} \left(\begin{array}{cc} &  A \\  A^\dagger & \end{array}\right) \right)
\exp\left(-i\frac{\theta}{2} \left(\begin{array}{cc} &  A^\dagger \\  A & \end{array}\right) \right)
,
\end{gather}
where both terms on the right-hand side can be implemented in the same way provided in the previous section.

\noindent
{\bf Plaquette:}
Consider the rotation of the plaquette term,
\begin{align}
R=\exp\left(-i\frac{\theta_P}{2}\left[P_{ABCD}+h.c.\right]\right)
,
\end{align}
with $P_{ABCD}=U_{AB}U_{BC}U^\dagger_{DC}U^\dagger_{AD}$.
For example, in the truncated matrix representation, the plaquette is 
\begin{gather}
\hskip -0.5cm
\resizebox{7.8cm}{!}{
$\displaystyle \left(\begin{array}{cccc}
 \0 & 1 & \0 & \0 \\
 \0 & \0 & 1 & \0 \\
 \0 & \0 & \0 & 1 \\
 \0 & \0 & \0 & \0 \\
\end{array}\right)
\!\!\!\otimes\!\!\!
\left(\begin{array}{cccc}
 \0 & 1 & \0 & \0 \\
 \0 & \0 & 1 & \0 \\
 \0 & \0 & \0 & 1 \\
 \0 & \0 & \0 & \0 \\
\end{array}\right)
\!\!\!\otimes\!\!\!
\left(\begin{array}{cccc}
 \0 & \0 & \0 & \0 \\
  1 & \0 & \0 & \0 \\
 \0 &  1 & \0 & \0 \\
 \0 & \0 & \ 1& \0 \\
\end{array}\right)
\!\!\!\otimes\!\!\!
\left(\begin{array}{cccc}
 \0 & \0 & \0 & \0 \\
  1 & \0 & \0 & \0 \\
 \0 &  1 & \0 & \0 \\
 \0 & \0 & \ 1& \0 \\
\end{array}\right)
$}.
\end{gather}
We first obtain the split
\begin{gather}
\exp\left(-i\frac{\theta_P}{2}\left[P_{ABCD}+h.c.\right]\right)
\approx \\ \nonumber 
\exp\left(-i\frac{\theta_P}{2}\left[P_{E}+h.c.\right]\right)
\exp\left(-i\frac{\theta_P}{2}\left[P_{O}+h.c.\right]\right)
\end{gather}
with 
\begin{gather}
\hskip -0.4cm
\resizebox{7.8cm}{!}{
$\displaystyle 
P_E=
\left(\begin{array}{cccc}
 \0 &  1 & \0 & \0 \\
 \0 & \0 & \0 & \0 \\
 \0 & \0 & \0 &  1 \\
 \0 & \0 & \0 & \0 \\
\end{array}\right)
\!\!\otimes\!\!
\left(\begin{array}{cccc}
 \0 & 1 & \0 & \0 \\
 \0 & \0 & 1 & \0 \\
 \0 & \0 & \0 & 1 \\
 \0 & \0 & \0 & \0 \\
\end{array}\right)
\!\!\otimes\!\!
\left(\begin{array}{cccc}
 \0 & \0 & \0 & \0 \\
  1 & \0 & \0 & \0 \\
 \0 &  1 & \0 & \0 \\
 \0 & \0 & \ 1& \0 \\
\end{array}\right)
\!\!\otimes\!\!
\left(\begin{array}{cccc}
 \0 & \0 & \0 & \0 \\
  1 & \0 & \0 & \0 \\
 \0 &  1 & \0 & \0 \\
 \0 & \0 & \ 1& \0 \\
\end{array}\right)
,
$}
\end{gather}
\begin{gather}
\hskip -0.4cm
\resizebox{7.8cm}{!}{
$\displaystyle 
P_O\!\!=\!\!
\left(\begin{array}{cccc}
 \0 & \0 & \0 & \0 \\
 \0 & \0 &  1 & \0 \\
 \0 & \0 & \0 & \0 \\
 \0 & \0 & \0 & \0 \\
\end{array}\right)
\!\!\otimes\!\!
\left(\begin{array}{cccc}
 \0 & 1 & \0 & \0 \\
 \0 & \0 & 1 & \0 \\
 \0 & \0 & \0 & 1 \\
 \0 & \0 & \0 & \0 \\
\end{array}\right)
\!\!\otimes\!\!
\left(\begin{array}{cccc}
 \0 & \0 & \0 & \0 \\
  1 & \0 & \0 & \0 \\
 \0 &  1 & \0 & \0 \\
 \0 & \0 & \ 1& \0 \\
\end{array}\right)
\!\!\otimes\!\!
\left(\begin{array}{cccc}
 \0 & \0 & \0 & \0 \\
  1 & \0 & \0 & \0 \\
 \0 &  1 & \0 & \0 \\
 \0 & \0 & \ 1& \0 \\
\end{array}\right)
.$}
\end{gather}
The two rotations can be achieved with ancilla, 
\begin{align*}
&\!\left(\! \begin{array}{c}  1 \\ 1 \end{array} \!\right)\!
\otimes
\exp\!\left(\!-i\frac{\theta_P}{2}\left[P_{E}+h.c.\right]\!\right)\!
\exp\!\left(\!-i\frac{\theta_P}{2}\left[P_{O}+h.c.\right]\!\right)\!
\Psi
\\&=
\exp\!\left(\!-i\frac{\theta}{2} \!\left(\!\begin{array}{cc} &  P_{E}+P_{E}^\dagger \\  P_{E}+P_{E}^\dagger & \end{array}\!\right)\! \!\right)\!
\\&\hspace{20pt}
\exp\!\left(\!-i\frac{\theta}{2} \!\left(\!\begin{array}{cc} &  P_{O}+P_{O}^\dagger \\  P_{O}+P_{O}^\dagger & \end{array}\!\right)\! \!\right)\!
\!\left(\! \begin{array}{c}  1 \\ 1 \end{array} \!\right)\! \otimes \Psi
.
\end{align*}
Notice that the commutation relations are still conserved during the split 
\begin{gather}
[\Pi,~U_E]=-U_E
,\quad
[\Pi,~U_O]=-U_O
.
\end{gather}

\noindent
{\bf Truncated ladder in Quantum Link Models:}
For Quantum Link Models in $U(1)$, if we put the matrix of $\Pi$ and $U$ directly into qubit, which is not usually what people do, we will get similar procedure as in the truncated ladder case above, where we can use the SVD to decompose the defective matrix into cyclic ladder matrix and diagonal matrix. In this respect, there is no big difference between the Quantum Link Models and the direct truncated one. 
Although, there is an extra burden from Quantum Link Models where we have to embed the matrix in the even dimension of $2^n$. This seems generally possible with universal gates. 

Since the construction of the unitary matrix $U$ up to large $N$ is known, we may conclude that for the choice of truncated ladder operators, mathematically we can go to large $N$ with no problem.

\section{Extension to non-Abelian gauge theory}
\label{app:ext}

It is useful to rephrase the treatment of the gauge field in the finite unitary matrix from a different angle in order to consider the possible extension to the non-Abelian theory.
Upon the electric basis, the electric and gauge link fields are represented as infinite dimensional matrices.
If we keep only the matrix elements with indices corresponding to the eigenvalues in the range $[\Pi_{\rm min},\Pi_{\rm max}]$, the commutation relation between $\Pi$ and $U$ will be conserved, since the electric field has only vanishing off-diagonal terms that have one index in the range and the other out.
In other words, after the truncation
\begin{align}
[\Pi,U']=- U'
,\quad{\rm and} \quad
e^{-i\alpha' \Pi} U' e^{i\alpha' \Pi} = e^{i\alpha' } U'
\end{align}
works for general real $\alpha'$.
The price is that the matrix representation of the gauge link field $U'$ is no longer unitary.
We wish to trade off the generality of $\alpha'$ for the unitary of $U'$.
In practice, we can complement $U'$ to a unitary matrix $U$ while still keeping the commutation relation to some degree, i.e.,
\begin{align}
U=U'+K
,\quad
e^{-i\alpha \Pi} U e^{-i\alpha \Pi} = e^{i\alpha}U
.
\label{eq:uk}
\end{align}
The complementation can be realized through the singular value decomposition, for example,
\begin{align}
U'={\mathscr V}
\left( \begin{tabular}{cc} ${\mathbb I}_{3\times 3}$ \\ & $0$ \end{tabular} \right)
{\mathscr W}^\dagger
,\quad
K={\mathscr V}
\left( \begin{tabular}{cc} $0_{3\times 3}$ \\ & $r$ \end{tabular} \right)
{\mathscr W}^\dagger
,
\end{align}
where $r$ is a general unitary matrix in the lower dimension but here it can only be 1.
Substituting the decomposition into Eq. (\ref{eq:uk}), we obtain $\alpha=2k\pi/4$ for integer $k$.
Thus we rediscover the finite representation of Weyl commutation relation.

\noindent
{\bf The (im)possible extension of unitary matrix to non-Abelian gauge theory:}
At the continuum level, the canonical quantisation of the non-Abelian gauge theory does not look very different from that of the Abelian theory.
There are similar non-trivial canonical commutation relations \cite{Jackiw:1988sf,Creutz:1978qg}
\begin{align}
[E^a_i(x),A^b_j(y)] = i \delta^{ab} \delta_{ij} \delta^3(x-y),
\end{align}
but for many copies with non-Abelian group indices which are denoted by $a,~b$.
We have similar $E_i^a=i\frac{\delta}{\delta A_i^a(x)}$ in the field eigen-basis.
However, on the lattice the two are very different.
For example, if we define the gauge link as in the Abelian theory,
\begin{align}
U_{\mu}(x)\!=\! \exp\left(-ig\ud xA_\mu(x)\right)
\!,\!~
A_{\mu}\!\equiv\! A_\mu^aT^a\!,~ T^a\!\!=\!\frac{\sigma^a}{2}
\!,\!
\end{align}
where for the example we have used $SU(2)$ group whose three generators can be represented by the Pauli matrices.
Due to the $SU(2)$ algebra, we can no longer obtain closed commutation relations, i.e., $[E,U]\not\propto U$.
The self-consistent commutation relations are first given by Kogut and Susskind  \cite{Kogut:1974ag} 
\begin{gather}
[(E_L)^a_{i}(x),U_j(y)]=\frac{\sigma^a}{2}U_i(x)\frac{g}{(\ud x)^2}\delta_{ij}\delta^3_{xy}
,\nonumber\\
[(E_R)^a_{i}(x),U_j(y)]=U_i(x)\frac{\sigma^a}{2}\frac{g}{(\ud x)^2}\delta_{ij}\delta^3_{xy}
,
,\nonumber\\
[(E_L)_i^a(x),(E_L)_j^b(y)]=-\frac{ig}{(\ud x)^2}\epsilon^{abc}(E_L)_i^c(x)\delta_{ij}\delta^3_{xy}
,\nonumber\\ \nonumber
[(E_R)_i^a(x),(E_R)_j^b(y)]=\frac{ig}{(\ud x)^2}\epsilon^{abc}(E_R)_i^c(x)\delta_{ij}\delta^3_{xy}
.
\end{gather}
Here we write explicitly the dependence on the lattice spacing $\ud x$.
Both $E_L$ and $E_R$ are the electric fields.
As we see, there exist more non-trivial commutation relations in the lattice theory, especially among the electric fields which do not exist in the continuum theory.
The apparent difference in the algebra can be dissolved under the continuum limit $\ud x\to 0$.
We take an expansion of the gauge link field, up to the next leading order of the lattice spacing $\ud x$,
\begin{align}
{\rm unitary}\quad
U_\mu(x) = 1-ig\ud x\frac{\sigma^a}{2}A_{\mu}^a(x)
+\cdots
.
\end{align}
The electric fields in the canonical quantisation must also admit an expansion on $\ud x$, for example,
\begin{align*}
(E_L)^a_i(x) &= \frac{i}{(\ud x)^3}\frac{\partial}{\partial A_i^a(x)}
+ \frac{g}{(\ud x)^2}\epsilon^{abc}A_i^b(x)\frac{\partial}{\partial A_i^c(x)}
+\cdots
\\
(E_R)^a_i(x) &= \frac{i}{(\ud x)^3}\frac{\partial}{\partial A_i^a(x)}
- \frac{g}{(\ud x)^2}\epsilon^{abc}A_i^b(x)\frac{\partial}{\partial A_i^c(x)}
+\cdots
\end{align*}
so the right continuum commutation relation is obtained in the limit $\ud x\to 0$.

At finite lattice spacing, we can adopt similar redefinition for the fields and obtain simple expressions,
\begin{align}
[E_L^a,E_L^b]=-i\epsilon^{abc}E_L^c
,\quad
[E_L^a,U]=\frac{\sigma^a}{2}U
,\\
[E_R^a,E_R^b]=i\epsilon^{abc}E_R^c
,\quad
[E_R^a,U]=U\frac{\sigma^a}{2}
.
\label{eq:E's}
\end{align}
The commutation relations between the left and right will vanish, $[E_L,E_R]=0$, but the left and right algebras must share the same Casimir, $E_L^aE_L^a=E_R^aE_R^a$, as a constraint.
Similar to the spin representation, the electric basis can be constructed with the eigenstates of three operators, $E^aE^a$, $E_{L}^3$ and $E_{R}^3$, satisfying \cite{Zohar:2012xf,Zohar:2014qma,DAndrea:2023qnr}
\begin{align}
E^aE^a|jm_Lm_R\rangle = j(j+1)|jm_Lm_R\rangle
,\nonumber\\
E^3_L|jm_Lm_R\rangle = m_L|jm_Lm_R\rangle
,\\ \nonumber
E^3_R|jm_Lm_R\rangle = m_R|jm_Lm_R\rangle
.
\end{align}
The representation is discrete but of infinite dimension, as the quantum number $j$ can go to infinity.
It is interesting to observe that there is no transition $j\to j',~j\neq j'$ happening to any $E$'s, and therefore the relevant matrix elements of $jj'$ must vanish.
As a result, if we adopt the truncation according to the Casimir operator $E^aE^a$, i.e. with all states $j<j_{\rm max}$, then all the commutation relations in Eq. (\ref{eq:E's})  will be exactly kept.  
The only violation occurs in $[U,U^\dagger]=0$, i.e. the unitarity of the gauge link.

On the other hand, there also exists Jordan-Schwinger boson representation of $SU(2)$ algebra, or the prepotential approach \cite{Mathur:2004kr}, where with independent boson operators $a$'s and $b$'s
\begin{gather}
E_L^a=\left(a_1^\dagger,a_2^\dagger\right)\frac{(\sigma^a)^T}{2}\left( \begin{array}{c} a_1 \\ a_2 \end{array} \right)
,~
E_R^a=\left(b_1^\dagger,b_2^\dagger\right)\frac{\sigma^a}{2}\left( \begin{array}{c} b_1 \\ b_2 \end{array} \right)
,\nonumber\\
U\!=\!
\frac{1}{\sqrt{N_a +1}}
\left( \begin{array}{cc} a_1^\dagger  &  -a_2  \\  a_2^\dagger & a_1  \end{array} \right)\!\!
\left( \begin{array}{cc} b_1^\dagger  &  b_2^\dagger  \\  -b_2 & b_1  \end{array} \right)
\frac{1}{\sqrt{N_b +1}}
.
\end{gather}
We note that the particle numbers, $N_a\equiv a^\dagger_1 a_1+ a^\dagger_2 a_2$ and $N_b\equiv b^\dagger_1 b_1+ b^\dagger_2 b_2$, in the denominators are crucial to keep the gauge link $U$ unitary.
Also, we can not find similar boson representation as above for $U(1)$ gauge theory, although there exists the boson representation in $U(1)$ Quantum Link Models
\cite{Chandrasekharan:1996ih,Kasper:2015cca,Felser:2019xyv,Magnifico:2020bqt}, where $U$ is not unitary.
Given $E^aE^a=\frac{N}{2}\left(\frac{N}{2}+1\right)$, the constraint on the Casimirs is now applied upon the particle numbers, $N_a=N_b$.
In the representation, the basis can simply be constructed from the particle numbers, $N$, $n_{a_1}=a^\dagger_1 a_1$ and $n_{b_1}=b^\dagger_1 b_1$, simply as $|N,n_{a_1}n_{b_1}\rangle $, which can further be used in the Loop-String-Hadron approach \cite{Raychowdhury:2019iki}.
The two bases above, the electric basis and the particle number basis, are in fact equivalent \cite{Davoudi:2020yln}.
Since the later is comparatively simpler without the Clebsch-Gordan coefficients, we are going to use the particle number basis in the following demonstration.

When truncating at $N_{\rm max}=1$, we obtain five eigenstates 
\begin{gather}
| 0,00 \rangle  = \left( \begin{tabular}{c} 1 \\ \0 \\ \0 \\ \0 \\ \0 \end{tabular} \right)
,\quad
| 1,00 \rangle  = \left( \begin{tabular}{c} \0 \\  1 \\ \0 \\ \0 \\ \0 \end{tabular} \right)
,\quad
| 1,01 \rangle  = \left( \begin{tabular}{c} \0 \\ \0 \\  1 \\ \0 \\ \0 \end{tabular} \right)
,\nonumber\\
| 1,10 \rangle  = \left( \begin{tabular}{c} \0 \\ \0 \\ \0 \\  1 \\ \0 \end{tabular} \right)
,\quad
| 1,11 \rangle  = \left( \begin{tabular}{c} \0 \\ \0 \\ \0 \\ \0 \\  1 \end{tabular} \right)
.
\end{gather}
The relevant matrices are
{\footnotesize
\begin{align}
\begin{aligned}[c]
E_L^1=
\frac{1}{2}\left(
\begin{tabular}{ccccc}
\0 & \0 & \0 & \0 & \0 \\
\0 & \0 & \0 &  1 & \0 \\
\0 & \0 & \0 & \0 &  1 \\
\0 &  1 & \0 & \0 & \0 \\
\0 & \0 &  1 & \0 & \0 
\end{tabular}
\right)
,~~
E_R^1=
\frac{1}{2}\left(
\begin{tabular}{ccccc}
\0 & \0 & \0 & \0 & \0 \\
\0 & \0 &  1 & \0 & \0 \\
\0 &  1 & \0 & \0 & \0 \\
\0 & \0 & \0 & \0 &  1 \\
\0 & \0 & \0 &  1 & \0 
\end{tabular}
\right)
\\
E_L^2=
\frac{1}{2}\left(
\begin{tabular}{ccccc}
\0 & \0 & \0 & \0 & \0 \\
\0 & \0 & \0 & -i & \0 \\
\0 & \0 & \0 & \0 & -i \\
\0 &  i & \0 & \0 & \0 \\
\0 & \0 &  i & \0 & \0 
\end{tabular}
\right)
,~~
E_R^2=
\frac{1}{2}\left(
\begin{tabular}{ccccc}
\0 & \0 & \0 & \0 & \0 \\
\0 & \0 &  i & \0 & \0 \\
\0 & -i & \0 & \0 & \0 \\
\0 & \0 & \0 & \0 &  i \\
\0 & \0 & \0 & -i & \0 
\end{tabular}
\right)
\\
E_L^3=
\frac{1}{2}\left(
\begin{tabular}{ccccc}
\0 & \0 & \0 & \0 & \0 \\
\0 & -1 & \0 & \0 & \0 \\
\0 & \0 & -1 & \0 & \0 \\
\0 & \0 & \0 &  1 & \0 \\
\0 & \0 & \0 & \0 &  1 
\end{tabular}
\right)
,~~
E_R^3=
\frac{1}{2}\left(
\begin{tabular}{ccccc}
\0 & \0 & \0 & \0 & \0 \\
\0 & -1 & \0 & \0 & \0 \\
\0 & \0 &  1 & \0 & \0 \\
\0 & \0 & \0 & -1 & \0 \\
\0 & \0 & \0 & \0 &  1 
\end{tabular}
\right)
\end{aligned}
,\\
\begin{aligned}[c]
U'=
\frac{1}{\sqrt{2}}
\left( \begin{tabular}{cc}
$\left( \begin{tabular}{ccccc}
\0 &  1 & \0 & \0 & \0  \\ 
\0\\
\0\\
\0\\
 1
\end{tabular} \right)
$&
$\left( \begin{tabular}{ccccc}
\0 & \0 & -1 & \0 & \0  \\ 
\0\\
\0\\
 1\\
\0
\end{tabular} \right)
$\\
$\left( \begin{tabular}{ccccc}
\0 & \0 & \0 & -1 & \0  \\ 
\0\\
 1\\
\0\\
\0
\end{tabular} \right)
$&
$\left( \begin{tabular}{ccccc}
\0 & \0 & \0 & \0 &  1  \\ 
 1\\
\0\\
\0\\
\0
\end{tabular} \right)
$\end{tabular} \right)
.
\end{aligned}
\end{align}
}%
It is easy to verify all the commutation relations in Eq. (\ref{eq:E's}) are exactly kept after the truncation. 
Therefore,
\begin{gather}
e^{i\alpha' E_L^a} U' e^{-i\alpha' E_L^a} = e^{i\alpha'\frac{\sigma^a}{2}}U' 
,\nonumber\\
e^{i\alpha' E_R^a} U' e^{-i\alpha' E_R^a} = U' e^{i\alpha'\frac{\sigma^a}{2}}
,
\end{gather}
work for any real $\alpha'$.
As inspired from the unitary gauge link approach in the $U(1)$ gauge theory, we  wish to ask whether we can complement the non-unitary $U'$ to a unitary one, while still keeping the relations above for some $\alpha$,
\begin{gather}
U=U'+K
,\quad
e^{i\alpha E_L^a} U e^{-i\alpha E_L^a} = e^{i\alpha\frac{\sigma^a}{2}}U
,\nonumber\\
e^{i\alpha E_R^a} U e^{-i\alpha E_R^a} = U e^{i\alpha\frac{\sigma^a}{2}}
.
\end{gather}
Generally, there is no solution of $K$ in the non-Abelian theory.
Notice via singular value decomposition
\begin{align}
U'\!=\!{\mathscr V}
\!\left(\! \begin{tabular}{cc} ${\mathbb I}_{4\times 4}$ \\ & $0_{6\times 6}$ \end{tabular} \!\right)\!
{\mathscr W}^\dagger
\!,\!~
K\!=\!{\mathscr V}
\!\left(\! \begin{tabular}{cc} $0_{4\times 4}$ \\ & $r_{6\times 6}$ \end{tabular} \!\right)\!
{\mathscr W}^\dagger
\!,\!
\end{align}
we can construct matrix $K$ from a general $6\times 6$ unitary matrix $r$.
The commutation relation requires
\begin{gather}
 e^{i\alpha E_L^a} K e^{-i\alpha E_L^a} = e^{i\alpha \frac{\sigma^a}{2}} K
\quad\Rightarrow \quad \nonumber \\
{\mathscr V}^\dagger e^{-i\alpha \frac{\sigma^a}{2}}e^{i\alpha E_L^a} {\mathscr V}
\left( \begin{tabular}{cc} 0 \\ & $r$ \end{tabular} \right)
 =  
\left( \begin{tabular}{cc} 0 \\ & $r$ \end{tabular} \right)
 {\mathscr W}^\dagger e^{i\alpha E_L^a}{\mathscr W}
,\\
 e^{i\alpha E_R^a} K e^{-i\alpha E_R^a} = K e^{i\alpha \frac{\sigma^a}{2}}
\quad\Rightarrow \quad \nonumber\\
{\mathscr V}^\dagger e^{i\alpha E_R^a} {\mathscr V}
\left( \begin{tabular}{cc} 0 \\ & $r$ \end{tabular} \right)
 =  
\left( \begin{tabular}{cc} 0 \\ & $r$ \end{tabular} \right)
 {\mathscr W}^\dagger e^{i\alpha \frac{\sigma^a}{2}}e^{i\alpha E_R^a}{\mathscr W}
\end{gather}
There only exists trivial solution $\alpha=4n\pi$.
Therefore, the unitary gauge link approach in QED can not be extended to the non-Abelian gauge theory.

\noindent
{\bf The extension of truncated ladder matrix to non-Abelian gauge theory:}
Generally, the electric fields working on the basis do not vary $N$ but the gauge-link fields increase or decrease $N$ by one.
So in the block of $N$'s, the electric fields are in diagonal, while the gauge-link fields are ladder operators, such as
\begin{center}
\includegraphics[height=3cm]{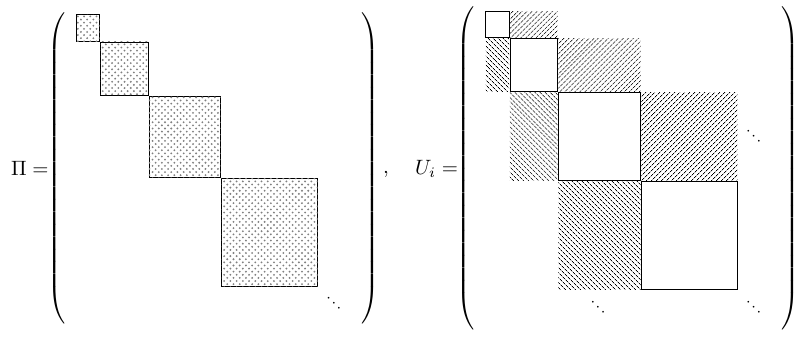}
\end{center}
In this sense, if we split the gauge-link fields according to both upper/lower and even/odd, each part will conserve the commutation relations with the electric fields, as the electric fields are in diagonal.
\begin{center}
\includegraphics[width=\linewidth]{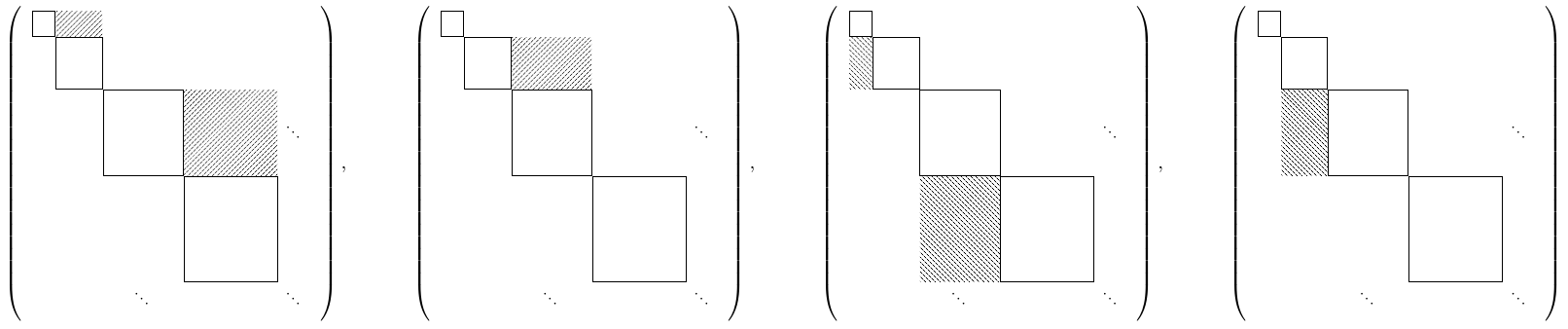}
\end{center}
More importantly, the square of each term vanishes.
Thus, we can split the plaquette into four terms, similar to the split in the $U(1)$ theory, and follow the same procedure there to implement the plaquette exactly.

\begin{figure*}[t!]
\includegraphics[width=\linewidth]{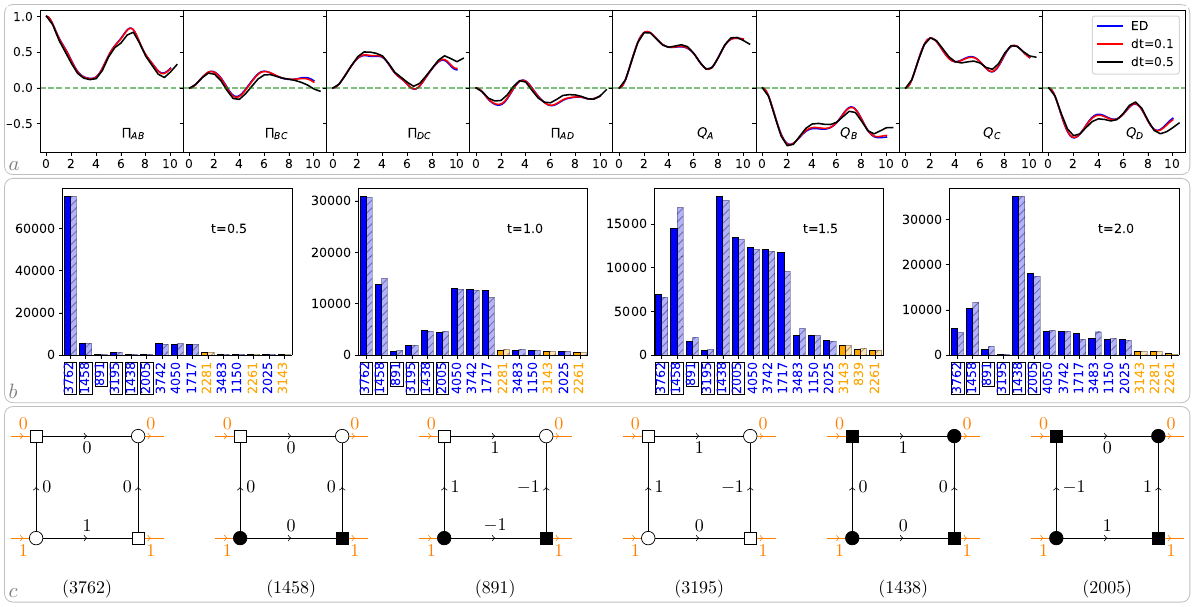}
\caption{%
(a) The evolution of the observables over time. Results from ED and Qiskit simulation with $\ud t=0.1$ and $0.5$ are plotted in blue, red and black individually. 
(b) The histogram of the states over time.
(c) An illustration of the dominating states.
}
\label{fig:dt}
\end{figure*}

\section{A study on $\ud t$}
\label{app:dt}

We have tested out different $\ud t$ with $n=2$ in 2D square case.
With 12+2 qubits, we are able to carry out qiskit simulation and even ED.
In the plot of the $\Pi$'s and $Q$'s over time, Fig. \ref{fig:dt}(a), the Trotter method agree with ED pretty well with $\ud t=0.1$, and even $\ud t=0.5$ follows the evolution closely among the time range.

We can also see from the histogram plot, Fig. \ref{fig:dt}(b), that $\ud t=0.1$ and $\ud t=0.5$ do not make a big difference at different time slices.
We expect when the constant external electric fields work on the vacuum, there will be pair production, whose result will screen the external fields.
That is, if we start with (3762), there will be states, such as (1458), happening during the evolution.
In the simulation, we set $m=0.1$, $g=\ud x=1$.
Since the fermion mass is relatively small, it is easier to excite fermion pairs than to excite higher electric fields.
This is why we get more state (1438) than state (891) in the statistics.

Since $\ud t =0.5$ provides good enough results and the number of Trotter steps is very limited, we are going to stick to $\ud t=0.5$ to probe longer time range for further tests.

\bibliographystyle{utphys} 
\bibliography{reference}

\end{document}